\documentclass[
aps,
prd,
floats,
floatfix,
twocolumn,
superscriptaddress,
nofootinbib,
noshowpacs]{revtex4-2}

\usepackage{hyperref}
\usepackage{amssymb}
\usepackage{amsmath}
\usepackage{physics}
\usepackage{latexsym}

\usepackage{verbatim}
\usepackage{mathrsfs}
\usepackage{amsfonts}

\usepackage{graphicx,subfigure}
\usepackage{epsfig}

\usepackage{overpic}
\usepackage[utf8]{inputenc}
\usepackage{rotating}
\usepackage{enumerate}

\usepackage{soul}

\usepackage[table]{xcolor}  
\usepackage{siunitx,booktabs}

\usepackage{xfrac}
\usepackage{ulem}
\usepackage{xcolor}
\usepackage{multirow}

\begin{document}

%%%%%%%%%%%
\title{Linear stability of nonrelativistic Proca stars}
%%%%%%%%%%%

%%%%%%%%%%%
\author{Emmanuel Ch\'avez Nambo}
\affiliation{Instituto de F\'isica y Matem\'aticas,
Universidad Michoacana de San Nicol\'as de Hidalgo,
Edificio C-3, Ciudad Universitaria, 58040 Morelia, Michoac\'an, M\'exico}
\author{Galo Diaz-Andrade}
\affiliation{Departamento de Física, División de Ciencias e Ingenierías, Campus León, Universidad de Guanajuato, 37150, León, México}
\author{Alberto Diez-Tejedor}
\affiliation{Departamento de Física, División de Ciencias e Ingenierías, Campus León, Universidad de Guanajuato, 37150, León, México}
\author{Edgar~Preciado-Govea}
\affiliation{Departamento de Física, División de Ciencias e Ingenierías, Campus León, Universidad de Guanajuato, 37150, León, México}
\author{Armando A. Roque}
\affiliation{Unidad Acad\'emica de F\'isica, Universidad Aut\'onoma de Zacatecas, 98060 Zacatecas, M\'exico}
\author{Olivier Sarbach}
\affiliation{Instituto de F\'isica y Matem\'aticas,
Universidad Michoacana de San Nicol\'as de Hidalgo,
Edificio C-3, Ciudad Universitaria, 58040 Morelia, Michoac\'an, M\'exico}
%%%%%%%%%%%

\date{\today}

%%%%%%%%%%%%%%%%%%%%%%%%%%%%%%%%%
\begin{abstract}
We study the linear stability of nonrelativistic Proca stars under generic perturbations. Using a combination of analytic and numerical methods, we demonstrate that, as expected, the ground state is always mode-stable. Additionally, we identify several mode-stable spherically symmetric excited states, including stationary states of constant and radial polarization, as well as multi-frequency states in case that the spin-spin self-interaction vanishes. The existence of stable excited states may have implications for spin-$1$ ultralight dark matter models.
\end{abstract}
%%%%%%%%%%%%%%%%%%%%%%%%%%%%%%%%%

\maketitle

%%%%%%%%%%%%%%%%%%%%%%%%%%%%%%%
\section{Introduction}
%%%%%%%%%%%%%%%%%%%%%%%%%%%%%%%

This paper continues the work initiated in Ref.~\cite{Nambo:2024hao}, where we studied the nonrelativistic classical equilibrium  configurations of a self-gravitating and self-interacting massive vector field, known as Proca stars (see Refs.~\cite{Jain:2021pnk, Zhang:2021xxa, Adshead:2021kvl, Amin:2022pzv, Jain:2022agt, Gorghetto:2022sue, Jain:2023tsr, Zhang:2023ktk, Chen:2024vgh, Zhang:2024bjo} for earlier work, and~\cite{Brito:2015pxa, SalazarLandea:2016bys, Brihaye:2017inn, Sanchis-Gual:2017bhw, Minamitsuji:2018kof, Sanchis-Gual:2018oui, Herdeiro:2020kba, Herdeiro:2020jzx, CalderonBustillo:2020fyi, Herdeiro:2021lwl, Sanchis-Gual:2022mkk, Rosa:2022tfv, Herdeiro:2023wqf, Wang:2023tly,  Sengo:2024pwk,Brito:2024biy,Lazarte:2025wlw} for their relativistic extension). Equilibrium configurations are typically classified  according to their stability properties, based on their ability to survive the presence of a generic small perturbation. Among them, stable equilibrium configurations are 
%specially relevant, 
of particular interest, since they are likely to appear at the end stage of a dynamical evolution. Therefore, differentiating between stable and unstable Proca stars is a relevant problem whose solution is the main purpose of this paper.

In Ref.~\cite{Nambo:2024hao}, we introduced the effective theory that describes the low energy  regime of a self-gravitating and self-interacting massive vector field. This theory is described by the $s=1$ Gross-Pitaevskii-Poisson system which depends on three parameters: the field's mass $m_0$ and the particle-particle $\lambda_n$ and spin-spin $\lambda_s$ self-interaction constants.\footnote{The parameter $\lambda_n$ is often referred to as the density-density self-interaction constant in the cold atom Bose-Einstein condensate literature~\cite{KAWAGUCHI2012253}. However, we prefer the term particle-particle self-interaction, since the spin-spin self-interaction is also generated by a density (the spin density in this case), and our terminology makes this parallel explicit.} Depending on the particular sector of the effective theory, we identified two different types of equilibrium configurations. In the {\it generic sector} of the theory, for which $\lambda_n$ is arbitrary and $\lambda_s \neq 0$, all equilibrium configurations are stationary states, and they are characterized by a wave function $\vec{\psi}$ that evolves harmonically in time. In contrast, in the {\it symmetry-enhanced sector}, where $\lambda_n$ is arbitrary and $\lambda_s=0$, the theory admits not only stationary states, but also multi-frequency states for which the components of the wave function evolve with two or three distinct frequencies~\cite{youtube2}.

When $\lambda_n$ and $\lambda_s$ lie in a certain range, there exists a ground state configuration whose energy is the lowest possible that can exist for a given fixed number of particles.\footnote{This is the ground state of the truncated dimension-6 effective theory presented in Eq.~(\ref{eq.action.nonrel}). While it does not necessarily coincide with the ground state of the underlying ultraviolet-complete theory, it still captures key features relevant to the stability of the nonrelativistic system. We discuss this further in Sec.~\ref{sec.ground}.} This state is described by a stationary state of constant polarization  whose wave function is spherical and has a positive, monotonically decreasing radial profile. Furthermore, the gravitational field of this state is the same as the one belonging to the ground state of the $s=0$ Gross-Pitaevskii-Poisson system. For most of this paper, we focus on the sector of the theory that admits a ground state. However, we will also discuss some instances in which equilibrium configurations exist, although the energy functional is unbounded from below.

When the energy functional is bounded  from below, in addition to the ground state, the $s=1$ Gross-Pitaevskii-Poisson system admits excited states, some of which are also spherically symmetric. These states include stationary states of constant polarization with higher energy than the ground state, stationary states of radial polarization, and multi-frequency states when $\lambda_s=0$. As indicated before, the key objective of this paper is to determine which of them are linearly stable.

Our analysis is based on a combination of analytic and numerical methods introduced by Harrison, Moroz and Tod in Ref.~\cite{2002math.ph...8045H}, which we have extended and adapted to study the  stability of equilibrium configurations of the $s=0$ (multi-field) Schr\"odinger-Poisson~\cite{Roque:2023sjl,Nambo:2023yut} and the $s=0$ Gross-Pitaevskii-Poisson~\cite{Nambo:2024gvs} systems under generic linear perturbations. Although strictly this approach only concerns the {\it mode-stability} of the system (i.e. its stability against a specific class of parametrized linear perturbations), we will refer to it simply as {\it stability} in this work. When applied to the $s=1$ Gross-Pitaevskii-Poisson systems, our analysis reveals  that the ground state configuration is always stable, as expected from its minimization property. Interestingly, however, our study also indicates that there exists excited states that are stable with respect to linear perturbations. This suggests that the vector  case includes new stable configurations that do not exist for a single scalar field~\cite{Moroz:1998dh, Moroz:1999, 2001PhLA..280..173T, 2002math.ph...8045H}, leading to a richer phenomenology for spin-1 ultralight dark matter models.

The structure of this paper is as follows: In Sec.~\ref{sec.EFT}, we summarize the main results of Ref.~\cite{Nambo:2024hao} and set the conventions for the reminder of this work. Then, in Sec.~\ref{sec.stationary}, we introduce the equations that describe the behavior of the linear perturbations around a general equilibrium configuration, and in Sec.~\ref{sec:linearized.system} we specialize on spherical equilibria. The main results of this paper are presented in Sec.~\ref{Sec:numerical}, where we classify spherical Proca stars according to their linear stability properties. A summary of this classification is given in Table~\ref{table.main}, where we highlight the stable configurations that are specific to the vector  case and have no counterpart in single scalar field theories. Finally, in Sec.~\ref{Sec:Conclussions} we conclude with a brief overview of the paper and discuss its relation to the  fully relativistic case. Technical details regarding the second variation of the energy functional which is used to prove the stability of some ground states, an equivalence between two perturbation systems, and the numerical implementation of the perturbation equations are presented in  appendixes. We work in natural units, for which $c=\hbar=1$.

%%%%%%%%%%%%%%%%%%%%%%%%%%%%%%%
\section{The \texorpdfstring{$s=1$}{s=1} Gross-Pitaevskii-Poisson system}\label{sec.EFT}
%%%%%%%%%%%%%%%%%%%%%%%%%%%%%%%

This section reviews the main results of Ref.~\cite{Nambo:2024hao} that are relevant for this paper. For a more detailed discussion, we invite the reader to consult the original work.

Our starting point is the nonrelativistic, low energy effective theory that describes a self-interacting  
vector field of mass $m_0$ coupled to Newtonian gravity, which is expressed in terms of the action\footnote{ Relativistic self-interacting Proca theories have been argued to exhibit pathologies~\cite{Clough:2022ygm, Mou:2022hqb, Coates:2022qia}, such as unstable modes and loss of hyperbolicity, although it has also been discussed~\cite{Barausse:2022rvg, Aoki:2022woy, Rubio:2024ryv} that these issues arise only outside the regime of validity of the effective theory. In the nonrelativistic limit considered here, the problematic high-frequency modes are suppressed, hence we expect the resulting Schr\"odinger-type theory to be well-posed. Furthermore, as shown in~\cite{Nambo:2024hao} it possesses an energy functional that is bounded from below for suitable values of $\lambda_n$ and $\lambda_s$.}
\begin{eqnarray}\label{eq.action.nonrel}
S[\mathcal{U}, \vec{\psi}] &=& \int dt\int dV \bigg[\frac{1}{8\pi G}\mathcal{U}\Delta\mathcal{U}-m_0\mathcal{U} n \\ 
&+&
\vec{\psi}^*\cdot\left(i\frac{\partial}{\partial t}+\frac{1}{2m_{0}}\Delta \right)\vec{\psi}-\frac{\lambda_n}{4m_{0}^{2}}n^{2}-\frac{\lambda_{s}}{4m_{0}^{2}}\vec{s}^{\,2}\bigg].\nonumber
\end{eqnarray}
This theory consists of a vector field $\vec{\psi}(t,\vec{x})$ and the Newtonian potential $\mathcal{U}(t,\vec{x})$, which are coupled through $-m_0\mathcal{U}n$. Here, $n:=\vec{\psi}^*\cdot\vec{\psi}$ is the particle number density and $\vec{s}:=-i\vec{\psi}^*\times\vec{\psi}$  the spin density, where the star denotes complex conjugation, and $\lambda_n$ and $\lambda_s$ are the particle-particle and spin-spin self-interaction constants, respectively. Equation~(\ref{eq.action.nonrel}) contains all local terms that are consistent with the Galilei group and are mass dimension 6 or lower. Higher-order operators are expected to be suppressed at small amplitudes $\vec{\psi}$ and are therefore not included in our effective theory.

Varying Eq.~(\ref{eq.action.nonrel}) with respect to $\vec{\psi}$ yields
\begin{subequations}\label{eq.GPPs=1}
\begin{align}\label{eq.GPs=1}
 i\frac{\partial\vec{\psi}}{\partial t} = -\frac{\Delta\vec{\psi}}{2m_0}+\frac{\lambda_n}{2m_0^2}n\vec{\psi}+i\frac{\lambda_s}{2m_0^2}\vec{s}\times\vec{\psi}+m_{0}\mathcal{U}\vec{\psi},
\end{align}
whereas the variation with respect to $\mathcal{U}$ leads to 
\begin{equation}\label{eq.Poisson}
\Delta\mathcal{U} = 4\pi G m_0n.
\end{equation}
\end{subequations}
Equations~(\ref{eq.GPPs=1}) define the $s=1$ Gross-Pitaevskii-Poisson system. In the absence of senfinteractions, $\lambda_n=\lambda_s=0$, this system reduces to the $s=1$ Schr\"odinger-Poisson system, which will be treated as a special case of the former.

By inverting the Laplacian in Eq.~(\ref{eq.Poisson}), one can eliminate the gravitational potential from~(\ref{eq.GPs=1}) 
and express this system as an integro-differential equation,
\begin{align}
\label{integro-differential_equation}
i\frac{\partial \vec{\psi}}{\partial t} &= \hat{{\mathcal{H}}} [\vec{\psi}] \vec{\psi},
\end{align}
with $\hat{{\mathcal{H}}} [\vec{\psi}]$ the $\vec{\psi}$-dependent Hamilton operator:
\begin{equation}
\label{eq.def.H}
\hat{\mathcal{H}}[\vec{\psi}] := -\frac{\Delta}{2m_0} + \frac{\lambda_n }{2m_0^2}n + i\frac{\lambda_s}{2m_0^2} \vec{s} \times + 4\pi Gm_0^2\Delta^{-1}(n).
\end{equation}
In this equation, $\Delta^{-1}$ denotes the inverse Laplacian, which acts on a generic function $f$ as follows:
\begin{equation}
[\Delta^{-1} (f)](\vec{x}) := -\frac{1}{4\pi}\int \frac{f(\vec{x}')}{|\vec{x} - \vec{x}'|}dV'.
\end{equation}

The $s=1$ Gross-Pitaevskii-Poisson system possesses several important quantities that are conserved in the time evolution and are useful for the characterization of the equilibrium solutions. For the purpose of this paper, we will be particularly interested in the energy, the total angular momentum, and the particle number, which are given by:
\begin{subequations}
\begin{eqnarray}
\mathcal{E}  &=& \int \left[\frac{|{\nabla}\vec{\psi}|^2}{2 m_0}+\frac{\lambda_n}{4m_0^2}n^{2}+\frac{\lambda_s}{4m_0^2}\vec{s}{\,}^2+\frac{m_{0}}{2}n\mathcal{U}\right]dV, \label{Eq.ConsEnergFunct}\quad\\
\vec{J} &=& - i\int (\vec{\psi}^{*}\times\vec{\psi}) dV + i\int [\vec{x}\times(\vec{\psi}^*\cdot\nabla \vec{\psi}) ] dV,\\
N &=& \int (\vec{\psi}^*\cdot\vec{\psi}) dV.
\end{eqnarray}
\end{subequations}
Positive (negative) values of $\lambda_n$ correspond to repulsive (attractive) particle-particle self-interactions, since the interaction energy increases (decreases) with the number density $n$. Conversely, positive (negative) values of $\lambda_s$ correspond to antiferromagnetic (ferromagnetic) spin-spin self-interactions: for $\lambda_s>0$ the energy is minimized by reducing the spin density's magnitude $|\vec{s}|$, whereas for $\lambda_s<0$ it is minimized by increasing $|\vec{s}|$. Note also that the total angular momentum $\vec{J}=\vec{S}+\vec{L}$ can be divided into the internal (or spin) angular momentum $\vec{S}$ and the orbital angular momentum $\vec{L}$, which are conserved separately. In addition, in absence of spin-spin self-interactions, the Hermitian second-rank tensor
\begin{equation}
\label{eq.globalQ}
\hat{Q}=\int (\vec{\psi}^*\otimes \vec{\psi}) dV
\end{equation}
is also conserved. This is a consequence of the accidental symmetry $\vec{\psi}(t,\vec{x})\mapsto \hat{U}\vec{\psi}(t,\vec{x})$ that appears when $\lambda_s=0$, where $\hat{U}$ is an arbitrary constant unitary $3\times 3$ matrix.

%%%%%%%%%%%%%%%%%%%%%
\subsection{Equilibrium configurations}
%%%%%%%%%%%%%%%%%%%%%
We define an equilibrium configuration as a critical point of the energy functional $\mathcal{E}[\vec{\psi}]$ with respect to variations of $\vec{\psi}$ that keep $N$ (if $\lambda_s\neq 0$) or $\hat{Q}$ (if $\lambda_s= 0$) fixed. They are characterized by the following ansatz: 
\begin{equation}
\label{eq.harmonic}
 \vec{\psi}(t,\vec{x}) = e^{-i \hat{E} t} \vec{\sigma}^{(0)}(\vec{x}) ,
\end{equation}
where $\vec{\sigma}^{(0)}$ is a complex vector-valued function of $\vec{x}$ and $\hat{E}$ is a Hermitian constant matrix. Equilibrium states have time-independent particle number density $n(t,\vec{x})=n(\vec{x})$ and, according to Poisson's equation, they give rise to static gravitational fields $\mathcal{U}(t,\vec{x})=\mathcal{U}(\vec{x})$. 

In practice, we distinguish between two different types of equilibrium configurations: {\it stationary states}, for which $\hat{E}$ is proportional to the identity matrix, and {\it multi-frequency states}, for which $\hat{E}$ is a constant linear transformation that cannot be expressed in the form $\hat{E}=EI$, where $E$ is a real constant and $I$ the $3\times3$ identity matrix. Whereas stationary states can exist for any  value of $\lambda_s$, multi-frequency states are only possible in the symmetry-enhanced sector of the theory, for which $\lambda_s=0$.

Introducing Eq.~(\ref{eq.harmonic}) into the integro-differential equation~(\ref{integro-differential_equation}) yields the nonlinear eigenvalue problem:
\begin{equation}\label{Eq:NLsystem}
\hat{E}\vec{\sigma}^{(0)} =\hat{\mathcal{H}}[ \vec{\sigma}^{(0)}]\vec{\sigma}^{(0)},
\end{equation}
where $\hat{\mathcal{H}}[ \vec{\sigma}^{(0)}]$ is given by Eq.~(\ref{eq.def.H}) replacing $\vec{\psi}$ by $\vec{\sigma}^{(0)}$.
This equation determines the equilibrium configurations of the $s=1$ Gross-Pitaevskii-Poisson system.

%%%%%%%%%%%%%%%%%%%%%%%%
\subsection{Spherical equilibrium configurations}
%%%%%%%%%%%%%%%%%%%%%%%%
In this paper we are particularly interested in spherical configurations, which are defined as those states that are invariant under an appropriate action of the $SO(3)$ group.\footnote{ Depending on how this action is chosen, $\vec{\psi}(t,\vec{x})\mapsto R\vec{\psi}(t, R^{-1}\vec{x})$ or $\vec{\psi}(t,\vec{x})\mapsto \vec{\psi}(t, R^{-1}\vec{x})$, different spherical configurations are obtained. In some instances, spherical symmetry is not manifest at the level of the vector field itself, which may single out a preferred spatial direction, even though other observables such as the gravitational potential remain spherically symmetric and depends only on $r$. We invite the reader to consult the original reference~\cite{Nambo:2024hao} for more details.}

Spherically symmetric stationary states can be expressed in the form:
\begin{equation}\label{eq.asph.stationary}
\vec{\psi}(t,\vec{x})= e^{-iE t}\sigma^{(0)}(r)\hat{\epsilon}(\vec{x}),
\end{equation}
where $\sigma^{(0)}$ is a real-valued function that depends only on the radial coordinate $r$, $E$ is a real constant, and $\hat{\epsilon}(\vec{x})$ is a polarization vector that, for convenience, we  normalize to one, $\hat{\epsilon}^*(\vec{x})\cdot\hat{\epsilon}(\vec{x})=1$. Up to symmetry transformations, there are three choices of the polarization vector that are compatible with spherical symmetry: $i)$~linear polarization, $\hat{\epsilon}(\vec{x})=\hat{e}_x$, $ii)$~circular polarization, $\hat{\epsilon}(\vec{x})=\hat{\epsilon}_z^{(+)}\equiv\frac{1}{\sqrt{2}}\left(\hat{e}_x+i\hat{e}_y\right)$, and $iii)$~radial polarization, $\hat{\epsilon}(\vec{x})=\hat{e}_r$. Note that for the first two cases the polarization is constant.\footnote{In the symmetry-enhanced sector, linear and circular polarizations are related to each other by a unitary transformation, and all constant polarization states, also known as elliptically polarized states, are equivalent. However, when $\lambda_s\neq 0$, this symmetry is broken, and it becomes necessary to distinguish between the linear and circular cases.}

If we introduce the ansatz~(\ref{eq.asph.stationary}) into the $s=1$ Gross-Pitaevskii-Poisson system~(\ref{eq.GPPs=1}), we obtain
\begin{widetext}
\begin{subequations}\label{s=1GPP.stationary}
\begin{eqnarray}
E\sigma^{(0)}&=&\left[-\frac{1}{2m_0}\left(\Delta_s-\frac{2\gamma}{r^2}\right)+\frac{\lambda_n+\alpha\lambda_s}{2m_0^2}\sigma^{(0)2} + m_0\mathcal{U}\right]\sigma^{(0)},\label{s=1GPP.stationary.1}\\
\Delta_s\mathcal{U} &=&4\pi Gm_0\sigma^{(0)2},\label{s=1GPP.stationary.2}
\end{eqnarray}
\end{subequations}
\end{widetext}
where $\Delta_s:=\frac{1}{r}\frac{d^2}{dr^2}r$ is the radial part of the Laplace operator and $\gamma$ and $\alpha$ are two parameters that depend on the polarization and take values: $i)$~$\gamma=0$, $\alpha=0$ if the polarization is linear, $ii)$~$\gamma=0$, $\alpha=1$ if the polarization is circular, and $iii)$~$\gamma=1$, $\alpha=0$ if the polarization is radial.

These equations must be solved with some appropriate boundary conditions. At the origin, the  regularity of the solutions demands
\begin{subequations}\label{Eq.BounCondr0}
\begin{align}
   \sigma^{(0)}(r=0)&=(1-\gamma)\sigma_{0},\quad \sigma^{(0)\prime}(r=0)=\gamma\sigma_0,\\
   \mathcal{U}(r=0)&=\mathcal{U}_0,\hspace{1.73cm} \mathcal{U}'(r=0)=0,
\end{align}
with $\sigma_0$ and $\mathcal{U}_0$ constants. Here and in the following, primes denote derivatives with respect to $r$. At infinity, we impose
\begin{equation}
\lim\limits_{r\to\infty}\sigma^{(0)}(r)=0, \quad \lim\limits_{r\to\infty}\mathcal{U}(r)=0,
\end{equation}
\end{subequations}
where the first condition is required for the solutions to have a finite total energy and the second is imposed without loss of generality.
Equations~(\ref{s=1GPP.stationary}) and~(\ref{Eq.BounCondr0}) define a nonlinear eigenvalue problem for the frequency $E$, where, for each value of $\sigma_0$, there exists an infinite, countable set of solutions $(E_n,\sigma^{(0)}_n)$ that can be labeled by the number of nodes $n$ of the wave function $\sigma_n^{(0)}(r)$ in the interval $0<r<\infty$, with $n=0,1,2,\ldots$. To the best of our knowledge, spherically symmetric constant polarization states were first presented almost simultaneously in Refs.~\cite{Jain:2021pnk, Zhang:2021xxa, Adshead:2021kvl, Amin:2022pzv} and were later extended to the relativistic regime in Ref.~\cite{Wang:2023tly}.

In the symmetry-enhanced sector ($\lambda_s=0$), in addition to the stationary states, there also exist multi-frequency states.
Spherically symmetric multi-frequency states
can be expressed in the form
\begin{equation}\label{eq.ansatz.multi.cartesian}
\vec{\psi}(t,\vec{x})= \sum_{j=1}^3 e^{-iE_j t}\sigma_j^{(0)}(r)\hat{e}_j,
\end{equation}
where $\sigma_j^{(0)}$ are three real-valued functions depending only on $r$, $E_j$ are three real numbers, and $\hat{e}_j$, $j=x,y,z$, denote the vectors constituting the Cartesian basis.

Introducing the ansatz~(\ref{eq.ansatz.multi.cartesian}) into the $s=1$ Gross-Pitaevskii-Poisson system~(\ref{eq.GPPs=1}) yields
\begin{subequations}\label{eqs.multi.frequency}
\begin{eqnarray}
E_i\sigma_i^{(0)}&=&\left[-\frac{1}{2m_0}\Delta_s+\frac{\lambda_n}{2m_0^2}\sum_{j}|\sigma_{j}^{(0)}|^2 + m_0\mathcal{U}\right]\sigma^{(0)}_i,\qquad \label{SEc2.2.2.2}\\
\Delta_s\mathcal{U} &=&4\pi Gm_0\sum_{j}|\sigma^{(0)}_{j}|^2.\label{s=1GPP.stationary.3}
\end{eqnarray}
\end{subequations}
These equations must be complemented with the following boundary conditions that guarantee that the solutions are regular at the origin and posses finite total energy,
\begin{subequations}\label{Eq.BounCond.multy}
\begin{align}
\sigma_j^{(0)}(r=0)&=\sigma_{j0},\quad \sigma_j^{(0)\prime}(r=0)=0,\\
\mathcal{U}(r=0)&=\mathcal{U}_{0},\hspace{0.75cm} \mathcal{U}'(r=0)=0,\\
\lim\limits_{r\to\infty}\sigma^{(0)}_j(r)&=0,\hspace{1cm} \lim\limits_{r\to\infty}\mathcal{U}(r)= 0,
\end{align}
\end{subequations}
where $\sigma_{j0}$ and $\mathcal{U}_0$ are constants. Equations~(\ref{eqs.multi.frequency}) and~(\ref{Eq.BounCond.multy}) define a nonlinear eigenvalue problem for the frequencies $E_j$. Similarly to the stationary case, for each $(\sigma_{x0},\sigma_{y0},\sigma_{z0})$, there exists an infinite, countable set of solutions $(E_{x,n_x}, E_{y,n_y}, E_{z,n_z}, \sigma_{x,n_x}^{(0)}(r),\sigma_{y,n_y}^{(0)}(r),\sigma_{z,n_z}^{(0)}(r))$ that can be labeled by the number of nodes $(n_x,n_y,n_z)$ of each wave function $\sigma_{x,n_x}^{(0)}(r)$, $\sigma_{y,n_y}^{(0)}(r)$, and $\sigma_{z,n_z}^{(0)}(r)$  in the interval $0<r<\infty$, with $n_n,n_y,n_z=0,1,2,\ldots$. As far as we are aware, multi-frequency states were first discussed in Ref.~\cite{Nambo:2024hao}.

The solutions to Eqs.~(\ref{s=1GPP.stationary}) and~(\ref{eqs.multi.frequency}) that satisfy the boundary conditions described above can be obtained using a shooting algorithm and were presented in Ref.~\cite{Nambo:2024hao}.

%%%%%%%%%%%%%%%%%%%%%%%%%%%
\subsection{Ground state}\label{sec.ground}
%%%%%%%%%%%%%%%%%%%%%%%%%%%
The ground state plays an important role in the study of field theories. Its existence is required for the consistency of the theory, and it is generally expected to be stable.

As shown in Ref.~\cite{Nambo:2024hao}, for fixed $N$, the energy~(\ref{Eq.ConsEnergFunct}) is bounded from below if and only if $\lambda_0\geq 0$, where
\begin{equation}
\lambda_0 := \left\{ \begin{array}{ll}
 \lambda_n, & \textrm{if } \lambda_s\geq 0,\\
 \lambda_n - |\lambda_s|, & {\textrm{if }}\lambda_s < 0.
\end{array} \right. \label{Eq:lambda0}
\end{equation}
Furthermore, the condition $\lambda_0\geq 0$ implies the existence of a ground state which is stationary and spherically symmetric, that is, as in Eq.~(\ref{eq.asph.stationary}), with $\sigma^{(0)}(r)$ a monotonically decreasing and positive function of $r$ (i.e. a function without nodes,  $n=0$). The polarization vector $\hat{\epsilon}(\vec{x})$ of this state is linear when $\lambda_s>0$, corresponding to the antiferromagnetic (also known as polar) phase of Proca stars which have vanishing spin density, and circular when $\lambda_s<0$, corresponding to the ferromagnetic phase, in which the spin density is maximal.\footnote{The {\it ferromagnetic} and {\it antiferromagnetic} phases are often described heuristically as having the individual particle spins aligned or misaligned, respectively. In a condensate, however, all particles occupy the same spinor state, so it is not meaningful to imagine the spin of different particles pointing in different directions. The correct characterization is in terms of the condensate spin density $\vec{s}$, which corresponds to the expectation value of the spin operator in the quantum theory. The antiferromagnetic phase has $\vec{s}=0$, while the ferromagnetic phase has maximal $|\vec{s}|$.} In the symmetry-enhanced sector ($\lambda_s=0$) all constant polarization states with the same node number are degenerate (that is, they are equivalent up to a symmetry transformation) and share the same energy.
For clarity, the region of parameter space in which a ground state exists is shown in Fig.~3 of Ref.~\cite{Nambo:2024hao}.

Stationary, spherically symmetric constant polarization states, including the ground state, are described by Eqs.~(\ref{s=1GPP.stationary}) with $\gamma=0$. These are the same equations that describe the stationary and spherically symmetric solutions of the $s=0$ Gross-Pitaevskii-Poisson system with $\lambda=\lambda_n+\alpha\lambda_s$, where $\lambda$ is the self-interaction constant associated with the scalar field~\cite{Nambo:2024gvs}. This implies that the gravitational field of the $s=1$ ground state is indistinguishable from that of the $s=0$ case (which is typically the only stable configuration in spherical symmetry within the single-field theory), and solely the existence of stable excited configurations in the vector field will allow the two systems to be distinguished through purely gravitational considerations.

The case $\lambda_0< 0$ is more involved and requires some clarification. Here, the energy~(\ref{Eq.ConsEnergFunct}) is unbounded from below and thus there does not exist a ground state. This represents a limitation of the truncated dimension-6 effective theory, which should be interpreted as a low energy approximation of a more fundamental theory. In the fundamental  theory, the higher-order self-interaction terms not included in Eq.~(\ref{eq.action.nonrel}) are expected to eventually produce a ground state and resolve the inconsistencies that arise in its truncated version.\footnote{In the context of effective theories, higher-order operators are expected for both $\lambda_0<0$ and $\lambda_0\ge 0$. The difference is that in the former case they are not only expected, they are also required for the existence of a ground state.} This also implies that our results are reliable only for a limited range of amplitudes  $0<\sigma_0<\sigma_0^{\textrm{crit}}$, for which the contribution of the higher-order operators remain negligible. The precise bound $\sigma_0^{\textrm{crit}}$ is determined by the ultraviolet completion of the theory and is thus model dependent. Nevertheless, as we shall see, the  truncated theory admits equilibrium configurations, some of which are stable and must persist after the completion.

%%%%%%%%%%%%%%%%%%%%%%%%%%%%%%%%%%%%%%
\section{Linearized equations}\label{sec.stationary}
%%%%%%%%%%%%%%%%%%%%%%%%%%%%%%%%%%%%%%
To analyze the stability of the equilibrium configurations, 
we follow the procedure presented in Refs.~\cite{Nambo:2023yut,Nambo:2024gvs} and study the behavior of small deviations of $\vec{\psi}(t,\vec{x})$ from the ansatz~(\ref{eq.harmonic}), which we parametrize as
\begin{equation}
\vec{\psi} (t,\vec{x})= e^{-i \hat{E} t}\left[\vec{\sigma}^{(0)}(\vec{x}) 
 + \epsilon\vec{\sigma}(t,\vec{x})+\mathcal{O}(\epsilon^2) \right].
\label{eq:ansatzPert}
\end{equation}
Here, $(\hat{E}, \vec{\sigma}^{(0)})$ is a solution of Eq.~(\ref{Eq:NLsystem}) and $\vec{\sigma}$ is a complex vector-valued function depending on $(t,\vec{x})$ that describes the perturbation to first order in the small parameter $\epsilon$. Recall that $\hat{E}$ is Hermitian, and further it is proportional to the identity when $\lambda_s\neq 0$.

Substituting the expansion~(\ref{eq:ansatzPert}) into Eq.~(\ref{integro-differential_equation}) and considering the first-order terms in $\epsilon$ we obtain the following evolution equation for $\vec{\sigma}$:
\begin{align}
\label{Eq:PerturbationEq}
    i \frac{\partial \vec{\sigma}}{\partial t} &= \left[\hat{\mathcal{H}}^{(0)} - \hat{E}\right] \vec{\sigma}\\ \nonumber
    &+ \hat{K}\left[\hbox{Re}\left(\vec{\sigma}^{(0)*} \cdot \vec{\sigma}\right)\right]\vec{\sigma}^{(0)} + i\frac{\lambda_s}{m_0^2} \hbox{Im}\left(\vec{\sigma}^{(0)*} \times \vec{\sigma}\right) \times \vec{\sigma}^{(0)},
\end{align}
with the linear (formally self-adjoint) operators $\hat{\mathcal{H}}^{(0)} := \hat{\mathcal{H}}[\vec{\sigma}^{(0)}]$ and 
\begin{align}
\label{Eq:Kdef}
\hat{K} &:= \frac{\lambda_n}{m_0^2} + 8\pi G m_0^2\Delta^{-1}.
\end{align}
Note that Eq.~(\ref{Eq:PerturbationEq}) is linear over the real numbers but not over the complex numbers (due to the last two terms on the right-hand side of this equation involving the real and imaginary parts). Consequently, the usual separation ansatz with time-dependence $e^{\lambda t}$ does not work unless $\lambda$ is real. However, as shown in Ref.~\cite{2002math.ph...8045H}, one can separate the time and space parts of $\vec{\sigma}$ by means of the following mode ansatz:
\begin{equation}
\label{Eq:PertAnsatz}
\vec{\sigma}(t,\vec{x}) = \left[ \vec{\mathcal{A}}(\vec{x})+\vec{\mathcal{B}}(\vec{x}) \right]e^{\lambda t} + \left[\vec{\mathcal{A}}(\vec{x})-\vec{\mathcal{B}}(\vec{x})\right]^{*}e^{\lambda^* t},
\end{equation}
which also works for complex values of $\lambda$. 
Here $\vec{\mathcal{A}}$ and $\vec{\mathcal{B}}$ are complex vector-valued functions depending only on $\vec{x}$ and $\lambda$ is a complex number. Substituting Eq.~(\ref{Eq:PertAnsatz}) into Eq.~(\ref{Eq:PerturbationEq}) and setting the coefficients in front of $e^{\lambda t}$ and $e^{\lambda^* t}$ to zero, one obtains\footnote{When $\lambda$ is real, $e^{\lambda t}$ and $e^{\lambda^* t}$ are linearly dependent. In this case one can assume without loss of generality that $\vec{\mathcal{A}}$ is real and $\vec{\mathcal{B}}$ is purely imaginary, and Eqs.~(\ref{Eq:equationAB}) remain correct. In the following, we shall adhere to this convention.}
\begin{subequations}
\label{Eq:equationAB}
    \begin{align}
        i \lambda \vec{\mathcal{A}} &= \left[\hat{\mathcal{H}}^{(0)} - \hat{E}\right] \vec{\mathcal{B}} + i \frac{\lambda_s}{2m_0^2} \vec{s}_0 \times (\vec{\mathcal{A}} - \vec{\mathcal{B}})\\ \nonumber
        &+ \frac{i}{2} \left\{\hat{K}\left[\vec{\sigma}^{(0)*} \cdot (\vec{\mathcal{A}} + \vec{\mathcal{B}}) + \vec{\sigma}^{(0)} \cdot (\vec{\mathcal{A}} - \vec{\mathcal{B}})\right]\right\} \hbox{Im} \; \vec{\sigma}^{(0)}\\ \nonumber
        &+ \frac{\lambda_s}{2m_0^2} \left[\vec{\sigma}^{(0)*} \times (\vec{\mathcal{A}} + \vec{\mathcal{B}}) - \vec{\sigma}^{(0)} \times (\vec{\mathcal{A}} - \vec{\mathcal{B}})\right] \times \hbox{Re} \;  \vec{\sigma}^{(0)},\\
        %%%%%%%%
        i \lambda \vec{\mathcal{B}} &= \left[\hat{\mathcal{H}}^{(0)} - \hat{E}\right] \vec{\mathcal{A}} - i \frac{\lambda_s}{2m_0^2} \vec{s}_0 \times (\vec{\mathcal{A}} - \vec{\mathcal{B}})\\ \nonumber
        &+ \frac{1}{2}\left\{\hat{K}\left[\vec{\sigma}^{(0)*} \cdot (\vec{\mathcal{A}} + \vec{\mathcal{B}}) + \vec{\sigma}^{(0)} \cdot (\vec{\mathcal{A}} - \vec{\mathcal{B}})\right]\right\} \hbox{Re} \; \vec{\sigma}^{(0)}\\ \nonumber
        &+ i\frac{\lambda_s}{2m_0^2} \left[\vec{\sigma}^{(0)*} \times (\vec{\mathcal{A}} + \vec{\mathcal{B}}) - \vec{\sigma}^{(0)} \times (\vec{\mathcal{A}} - \vec{\mathcal{B}})\right] \times \hbox{Im} \;\vec{\sigma}^{(0)},
    \end{align}
\end{subequations}
where $\vec{s}_0:=-i\vec{\sigma}^{(0)*}\times\vec{\sigma}^{(0)}$ is the spin density associated with the background solution. It is important to stress that in order to obtain Eqs.~(\ref{Eq:equationAB}) we have assumed that $\hat{E}$ is real-valued; hence, for multi-frequency states, these equations are only valid in the basis that diagonalizes the transformation $\hat{E}$.

Equations~(\ref{Eq:equationAB}) constitute a linear eigenvalue problem for the constant $\lambda$. The presence of an eigenvalue with positive real part $\lambda_R$ gives rise to an exponentially growing mode with associated lifetime $t_{\textrm{life}}\sim 1/\lambda_R$ and implies linear instability. If no such modes are present one calls the configuration mode-stable. In general, mode-stability does not  necessarily guarantee that the linearized system is stable unless one can prove that mode solutions of the form~(\ref{Eq:PertAnsatz}) are complete. Furthermore, the stability of the linearized system does not automatically imply the stability of the nonlinear system. Nonetheless, for $\lambda_s=0$, we have verified the conclusions of this work using a numerical evolution code, the results of which will be presented elsewhere. This provides additional support for the stability conditions that we identify in this paper, suggesting that they may hold even in the nonlinear regime.

%%%%%%%%%%%
\subsection{Basic properties of the linearized equations}
%%%%%%%%%%%
Next, we review some properties of the solutions to the system~(\ref{Eq:equationAB}) that are directly relevant for the analysis of the numerical results presented in Sec.~\ref{Sec:numerical}.

On the one hand, there always exists a zero-mode $\lambda=0$ solution of the form $( \vec{\mathcal{A}}, \vec{\mathcal{B}}) = \frac{1}{2}(-\beta\Im \vec{\sigma}^{(0)}, i\beta \Re\vec{\sigma}^{(0)})$, with $\beta$ an arbitrary real constant. This mode corresponds to a trivial ``perturbation'' that simply rotates the function $\vec{\sigma}^{(0)}(\vec{x})$ by an angle $\beta$ into the complex plane, yielding a new configuration $\psi(t, \vec{x})$ that is physically indistinguishable from the one in Eq.~(\ref{eq.harmonic}).  Recognizing this fact will help us identifying spurious modes in our numerical results, i.e. modes that appear to be unstable due to a small non-vanishing real part of $\lambda$ originating from numerical error.

On the other hand, for configurations for which $\vec{\sigma}^{(0)}$ is real-valued (i.e. multi-frequency states and stationary states of linear or radial polarization), the solutions always come in quadruples: if $(\lambda, \vec{\mathcal{A}}, \vec{\mathcal{B}})$ satisfies Eqs.~(\ref{Eq:equationAB}), then so do $(-\lambda, \vec{\mathcal{A}}, -\vec{\mathcal{B}})$, $(\lambda^*, \vec{\mathcal{A}}^*,-\vec{\mathcal{B}}^*)$ and $(-\lambda^*, \vec{\mathcal{A}}^*,\vec{\mathcal{B}}^*)$. In principle, this property does not apply to stationary states of circular polarization, which are described by complex-valued functions $\vec{\sigma}^{(0)}$ and for which we have not been able to identify a corresponding pairing of the eigenvalues $\lambda$. However, when $\lambda_s=0$, constant polarization states become degenerate, and there always exist a constant unitary transformation that connects linear and circular polarizations. Consequently, in this case, stationary circular polarization states exhibit the same quadruple structure  $\{\lambda, -\lambda, \lambda^*, -\lambda^* \}$ as linear polarization states, although this is not evident from Eqs.~(\ref{Eq:equationAB}).
In the remainder of this work, we will count unstable modes according to the number of eigenvalues with positive real parts. Thus, when the solutions appear in quadruples and both the real and imaginary parts of $\lambda$ are nonzero, the pair of eigenvalues $\lambda$ and $\lambda^*$ having positive real parts will be counted as two unstable modes, given that they correspond to two linearly independent eigenfunctions.

Finally, in App.~\ref{Sec:SecondVariation}, using the second variation of the energy functional~(\ref{Eq.ConsEnergFunct}),  we provide an analytic proof for the linear stability of the ground state in the case for which $\lambda_0\ge 0$ and $\lambda_s\ge 0$ (upper shaded region of Fig.~3 in~\cite{Nambo:2024hao}). To stablish the linear stability of the ground state when $\lambda_0\ge0$ and $\lambda_s<0$ (lower shaded region of the same figure), and to classify  excited spherical Proca stars according to their stability properties, we will employ a combination of analytic and numerical methods, which we describe in  Sec~\ref{Sec:numerical}.

%%%%%%%%%%%
\section{Linearized systems for spherical equilibria}
\label{sec:linearized.system}
%%%%%%%%%%%

In the previous section, we presented the equations that describe the evolution of a general linear perturbation around an arbitrary equilibrium configuration of the $s=1$ Gross-Pitaevskii-Poisson system. Here, we concentrate specifically on spherical equilibrium configurations. This allows us to decouple the linearized equations~(\ref{Eq:equationAB}) into a family of purely radial systems by expanding the perturbations in terms of (scalar or vector) spherical harmonics.

The main result of this section is that for stationary states of linear, circular, and radial polarization, as well as for multi-frequency states, the linearized system can be cast into the following general schematic form:
\begin{equation}\label{eq.general system}
i\lambda \left( \begin{array}{c}
X_{JM} \\ Y_{JM} \\ Z_{JM} \end{array} \right) = \left( \begin{array}{ccc}
 M_{JM}^{11} & M_{JM}^{12} & M_{JM}^{13}  \\
 M_{JM}^{21} & M_{JM}^{22} & M_{JM}^{23}  \\ 
 M_{JM}^{31} & M_{JM}^{32} & M_{JM}^{33}  \\
\end{array} \right)\left( \begin{array}{c}
X_{JM} \\ Y_{JM} \\ Z_{JM} \\ \end{array} \right),
\end{equation}
where $J$ refers to the angular momentum number of the perturbation and $M$ to the associated magnetic quantum number and assume values $J=0,1,2,\ldots$ and $M=-J,-(J-1), \ldots, J$. The particular realization of the variables $X_{JM}$, $Y_{JM}$, and $Z_{JM}$ and the matrix $M_{JM}$ (which is a function of the background equilibrium configuration) depend on the case of interest, as we discuss next.

Before doing so, we introduce a set of dimensionless quantities that simplify the equations and facilitate the presentation of our numerical results in the next section:
\begin{subequations}\label{eq.code.numbers1}
\begin{eqnarray}
{\displaystyle t:=\frac{4\pi G m_{0}^3 }{\lambda_* }t^{phys},} &\quad {\displaystyle  \vec{x} := \frac{\sqrt{8\pi G} m_{0}^2}{\lambda_*^{1/2}} \vec{x}^{phys},} \\
{\displaystyle \mathcal{U} := \frac{\lambda_* }{4\pi G m_0^2}\mathcal{U}^{phys},} &\quad {\displaystyle \vec{\psi}:= \frac{\lambda_* }{\sqrt{8\pi G}m_0^{5/2}}\vec{\psi}^{phys},} \\
{\displaystyle \lambda_n := \frac{\lambda_n^{phys}}{\lambda_*},} 
&\quad {\displaystyle  \lambda_s := \frac{\lambda_s^{phys}}{\lambda_*}},
\end{eqnarray}
\end{subequations}	
where $\lambda_*>0$ is a characteristic self-interaction scale that we can choose at our convenience. From this point onwards, $t$, $\vec{x}$, $\mathcal{U}$, $\ldots$ denote dimensionless variables, and we will use the superscript {\it phys} when we need to denote a dimensionfull quantity. 

%%%%%%%%%%%%%%%%%%%%%%%%
\begin{widetext}
\subsubsection{Stationary states of constant linear polarization}\label{sec.linear.pol}
%%%%%%%%%%%%%%%%%%%%%%%%
Since in this case the background solution $\vec{\sigma}^{(0)}(\vec{x})$ is real-valued, Eqs.~(\ref{Eq:equationAB}) reduce to
\begin{subequations}\label{Eq:PerturbationSigmaReal}
\begin{align}
    i \lambda \vec{\mathcal{A}} &= \left[\hat{\mathcal{H}}^{(0)} - E\right] \vec{\mathcal{B}} + 2\lambda_s \left( \vec{\sigma}^{(0)} \times \vec{\mathcal{B}}\right) \times \vec{\sigma}^{(0)},\\
    i \lambda \vec{\mathcal{B}} &= \left[\hat{\mathcal{H}}^{(0)} - E\right] \vec{\mathcal{A}} + 2\hat{K} \left( \vec{\sigma}^{(0)} \cdot \vec{\mathcal{A}}\right) \vec{\sigma}^{(0)},
\end{align}
\end{subequations}
where we have used the fact that for linearly polarized Proca stars $\vec{s}_{0}$ vanishes and $\hat{E}=EI$, with $I$ the unit matrix. In dimensionless units the operators $\hat{\mathcal{H}}^{(0)}$ and $\hat{K}$ read
\begin{subequations}
\begin{eqnarray}
\hat{\mathcal{H}}^{(0)} &=& -\Delta+\lambda_n n_0 +i\lambda_s \vec{s}_0 \times +\Delta^{-1}(n_0),\\
\hat{K} &=& \lambda_n+\Delta^{-1},
\end{eqnarray}
\end{subequations}
with $n_0=\vec{\sigma}^{(0)*}\cdot\vec{\sigma}^{(0)}$ and $\vec{s}_0=-i\vec{\sigma}^{(0)*}\times\vec{\sigma}^{(0)}$.

To proceed, we expand the fields $\vec{\mathcal{A}}(\vec{x})$ and $\vec{\mathcal{B}}(\vec{x})$ in the following way
\begin{subequations}\label{Eq:CartesianPert}
\begin{eqnarray}
\vec{\mathcal{A}}(\vec{x}) &=& \sum_i\sum_{JM} A_{JM}^i(r)Y^{JM}(\vartheta,\varphi)\hat{e}_i ,\\
\vec{\mathcal{B}}(\vec{x}) &=& \sum_i\sum_{JM} B_{JM}^i(r)Y^{JM}(\vartheta,\varphi)\hat{e}_i ,
\end{eqnarray}
\end{subequations}
where $A^i_{JM}$ and $B^i_{JM}$ ($i=x,y,z$) are complex-valued functions depending only on the radial coordinate and $Y^{JM}$ are the scalar spherical harmonics. Substituting the decomposition~(\ref{Eq:CartesianPert}) into Eqs.~(\ref{Eq:PerturbationSigmaReal}), we arrive at a system of the form~(\ref{eq.general system}), where
\begin{equation}\label{eq.defXYZ1}
X_{JM} = \left( \begin{array}{c}
A_{JM}^x \\ B_{JM}^x \end{array} \right), \quad 
Y_{JM} = \left( \begin{array}{c}
A_{JM}^y \\ B_{JM}^y \end{array} \right), \quad 
Z_{JM} = \left( \begin{array}{c}
A_{JM}^z \\ B_{JM}^z \end{array} \right),
\end{equation}
and
\begin{subequations}\label{eq.matriz.linear}
\begin{eqnarray}
 M_{JM}^{11} &=& \left( \begin{array}{cc}
 0 & \hat{\mathcal{H}}^{(0)}_J-E \\
\hat{\mathcal{H}}^{(0)}_J+2\sigma^{(0)}\hat{K}_J\left[\sigma^{(0)}\right]-E & 0  
\end{array} \right),
\\
M_{JM}^{22} &=&  \left( \begin{array}{cc}
 0 & \hat{\mathcal{H}}^{(0)}_J+2\lambda_s\sigma^{(0)2}-E \\
 \hat{\mathcal{H}}^{(0)}_J-E & 0  
\end{array} \right),\label{eq.matriz.linear2}  
\\
M_{JM}^{33} &=& M_{JM}^{22}, 
\end{eqnarray}
\end{subequations}
the other elements of the matrix  $M_{JM}$ being zero. In these expressions, the operators $\mathcal{\hat H}_J^{(0)}$ and $\hat{K}_J$ are defined by
\begin{subequations}
\begin{align}
\hat{\mathcal{H}}_J^{(0)}&:=-\Delta_J +\lambda_n\sigma^{(0)2} + \Delta_s^{-1}(\sigma^{(0)2}),\label{Eq.OpeH0_j}\\
\hat{K}_J &:=\lambda_n+\Delta_J^{-1},
\label{eq:H0LDef}
\end{align}
where %$\triangle_J := \triangle_s - J(J+1)/r^2$,
\begin{eqnarray}
\Delta_J &:=& \Delta_s - \frac{J(J+1)}{r^2},\\
\Delta_J^{-1}(f)(r) &:=& -\frac{1}{2J+1}\int\limits_0^\infty \frac{r_{<}^{J}}{r_{>}^{J+1}} f(\tilde{r})\tilde{r}^2 d\tilde{r},
\label{Eq:LapJInv}
\end{eqnarray}
\end{subequations}
with $r_<:=\min\{ r,\tilde{r} \}$ and $r_>:=\max\{ r,\tilde{r} \}$. The action of the operator $\hat{K}_J\left[\sigma^{(0)}\right]$ on a function $A$ is defined to be $\hat{K}_J\left[\sigma^{(0)}\right] A := \hat{K}_J\left[\sigma^{(0)}A \right]$. Note that $\Delta_s:=\Delta_{J=0}$, $\Delta_s^{-1}:=\Delta_{J=0}^{-1}$, and $\hat{\mathcal{H}}_J^{(0)}$ contains only the part of the Hamiltonian that is independent of the spin-spin self-interaction, that in any case vanishes when the polarization is linear. Finally, we stress that, at this point, the scale $\lambda_*$ appearing in the definition of the dimensionless variables is still free.

Note that the equation $i\lambda X_{JM}=M_{JM}^{11}X_{JM}$ coincides with Eqs.~(35) in Ref.~\cite{Nambo:2024gvs} for the system that describes the behavior of generic linear perturbations of nonrelativistic boson stars. However, this equation is now only a part of the whole system~(\ref{eq.general system}), due to the fact that we are dealing with a vector instead of a scalar field.

%%%%%%%%%%%%%%%%%%%%%%%%
\subsubsection{Stationary states of constant circular polarization}\label{sec.circular.pol}
%%%%%%%%%%%%%%%%%%%%%%%%
In this case the background solution $\vec{\sigma}^{(0)}(\vec{x})$ is complex-valued and $\vec{s}_0\neq \vec{0}$, so no drastic simplifications of Eqs.~(\ref{Eq:equationAB}) occur like in the previous case. Substituting Eqs.~(\ref{Eq:CartesianPert}) into Eqs.~(\ref{Eq:equationAB}) we again obtain, after some algebra,  a system of the form~(\ref{eq.general system}), where $X_{JM}$, $Y_{JM}$, and $Z_{JM}$ are given by Eq.~(\ref{eq.defXYZ1}) and 
\begin{subequations}\label{eq.matriz.circular}
\begin{eqnarray}
& M_{JM}^{11}= \left(\begin{array}{cc} 0 & \hat{\mathcal{H}}^{(0)}_J-E \\
\hat{\mathcal{H}}^{(0)}_J+\lambda_s\sigma^{(0)2}+\sigma^{(0)}\hat{K}_J\left[\sigma^{(0)}\right]-E & 0  
\end{array}\right),\quad 
M_{JM}^{12}= - i\sigma^{(0)} 
\left( \begin{array}{cc}
 \lambda_s\sigma^{(0)} & 0 \\
 0 & 2\lambda_s\sigma^{(0)} +\hat{K}_J\left[\sigma^{(0)}\right] 
\end{array} \right),\qquad &
\\
& M_{JM}^{21}= i\sigma^{(0)} 
\left( \begin{array}{cc}
 2\lambda_s\sigma^{(0)}+\hat{K}_J\left[\sigma^{(0)}\right] & 0 \\
 0 & \lambda_s\sigma^{(0)}  
\end{array} \right),\quad 
M_{JM}^{22}= (M_{JM}^{11})^T, &
\\
& M_{JM}^{33}=
 \left( \begin{array}{cc}
 0 & \hat{\mathcal{H}}^{(0)}_J+\lambda_s\sigma^{(0)2}-E \\
 \hat{\mathcal{H}}^{(0)}_J+\lambda_s\sigma^{(0)2}-E & 0  
\end{array} \right), &
\end{eqnarray}
\end{subequations}
with $M_{JM}^{13}=M_{JM}^{23}=M_{JM}^{31}=M_{JM}^{32}=0$ and the superscript $T$ denoting the transposed matrix. 

In the absence of spin-spin self-interactions, all constant polarization states are related to each other by a global unitary transformation. Consequently, when $\lambda_s=0$, the equations governing the linear perturbations of a stationary state of constant circular polarization must be equivalent to those of a stationary state of constant linear polarization. We prove the equivalence between these two systems in App.~\ref{Sec:equivalence}.

%%%%%%%%%%%%%%%%%%%%%%%%
\subsubsection{Stationary states of radial polarization}
%%%%%%%%%%%%%%%%%%%%%%%%
As in the linear case, the background solution $\vec{\sigma}^{(0)}(\vec{x})$ associated with a radially polarized Proca star is real-valued, so we can use again the reduced system~(\ref{Eq:PerturbationSigmaReal}). In this case, as explained in Ref.~\cite{Nambo:2023yut}, one can decouple the equations by expanding the perturbations in terms of vector spherical harmonics \cite{Khersonskii:1988krb}
\begin{subequations}\label{Eq:RadialPert}
\begin{eqnarray}
    \vec{\mathcal{A}}(\vec{x}) &=& \sum_{JM} \left[A_{JM}^r(r) \vec{Y}^{JM}(\vartheta,\varphi) + A_{JM}^{(1)}(r) \vec{\Psi}^{JM}(\vartheta,\varphi) + A_{JM}^{(2)}(r) \vec{\Phi}^{JM}(\vartheta,\varphi)\right],\\
     \vec{\mathcal{B}}(\vec{x}) &=& \sum_{JM} \left[B_{JM}^r(r) \vec{Y}^{JM}(\vartheta,\varphi) + B_{JM}^{(1)}(r) \vec{\Psi}^{JM}(\vartheta,\varphi) + B_{JM}^{(2)}(r) \vec{\Phi}^{JM}(\vartheta,\varphi)\right].
\end{eqnarray}
\end{subequations}
Substituting the decomposition~(\ref{Eq:RadialPert}) into Eqs.~(\ref{Eq:PerturbationSigmaReal}), we obtain a system of the form~(\ref{eq.general system}), where now
\begin{equation} \label{eq.defXYZ2}
X_{JM} = \left( \begin{array}{c}
A_{JM}^{r} \\%[.1cm] 
B_{JM}^{r} \end{array} \right), \quad 
Y_{JM} = \left( \begin{array}{c}
A_{JM}^{(1)} \\%[.1cm] 
B_{JM}^{(1)} \end{array} \right), \quad 
Z_{JM} = \left( \begin{array}{c}
A_{JM}^{(2)} \\%[.1cm] 
B_{JM}^{(2)} \end{array} \right),
\end{equation}
and
\begin{subequations}\label{eq.matriz.radial}
\begin{eqnarray}
& M_{JM}^{11} = \left(\begin{array}{c c c}
 0 & & \mathcal{\hat H}_J^{(0)} +\displaystyle{\frac{2}{r^2}} - E \\
\mathcal{\hat H}_J^{(0)} +\displaystyle{\frac{2}{r^2}} +2\sigma^{(0)}\hat{K}_J\left[\sigma^{(0)}\right]- E & & 0
\end{array} \right), \quad  M_{JM}^{12} = -\displaystyle{\frac{2}{r^2}}\left(\begin{array}{c c c}
 0 & & \sqrt{J(J+1)}\\
 \sqrt{J(J+1)}& & 0
\end{array} \right),\;\;&\\
& M_{JM}^{21} = M_{JM}^{12}, \quad M_{JM}^{22} = \left( \begin{array}{ccc}
 0 & & \mathcal{\hat H}_J^{(0)} + 2\lambda_s \sigma^{(0)2} -E \\
 \mathcal{\hat H}_J^{(0)} - E & & 0 
\end{array} \right),& \\
&M_{JM}^{33}=M_{JM}^{22},&
\end{eqnarray}
\end{subequations}
with $M_{JM}^{13}=M_{JM}^{23}=M_{JM}^{31}=M_{JM}^{32}=0$. In absence of self-interactions ($\lambda_n=\lambda_s=0$) this system reduces to Eqs.~(41) in Ref.~\cite{Nambo:2023yut} for $\ell=1$ boson stars.

%%%%%%%%%%%%%%%%%%%%%%%%
\subsubsection{Multi-frequency states}
%%%%%%%%%%%%%%%%%%%%%%%%
Multi-frequency states  exist only in the symmetry-enhanced sector, for which $\lambda_s=0$. In addition, for these states the background solution $\vec{\sigma}^{(0)}(\vec{x})$ is real-valued and Eqs.~(\ref{Eq:equationAB}) then simplify to
\begin{subequations}\label{Eq:Perturbation.multi}
\begin{align}
    i \lambda \vec{\mathcal{A}} &= \left[\hat{\mathcal{H}}^{(0)} - \hat{E}\right] \vec{\mathcal{B}} ,\\
    i \lambda \vec{\mathcal{B}} &= \left[\hat{\mathcal{H}}^{(0)} - \hat{E}\right] \vec{\mathcal{A}} + 2\hat{K} \left( \vec{\sigma}^{(0)} \cdot \vec{\mathcal{A}}\right) \vec{\sigma}^{(0)}.
\end{align}
\end{subequations}
To proceed, we expand the fields $\vec{\mathcal{A}}(\vec{x})$ and $\vec{\mathcal{B}}(\vec{x})$ as in Eqs.~(\ref{Eq:CartesianPert}). If we substitute this decomposition into Eqs.~(\ref{Eq:Perturbation.multi}), we obtain, after some algebra, a system of equations that can be expressed in the form~(\ref{eq.general system}), where $X_{JM}$, $Y_{JM}$, and $Z_{JM}$ are given by Eq.~(\ref{eq.defXYZ1}), and\footnote{Note that in the limit for which $\sigma^{(0)}_y=\sigma^{(0)}_z=0$,  a multi-frequency state reduces to a stationary state with linear polarization and $\sigma^{(0)}=\sigma^{(0)}_x$. It is simple to verify that in this limit, if we choose $E_x=E_y=E_z=E$, the matrices $M_{JM}^{ij}$ in Eq.~(\ref{Eq:MijJMMulti}) reduce to those in Eqs.~(\ref{eq.matriz.linear}) with $\lambda_s=0$.}
\begin{subequations}\label{Eq:MijJMMulti}
\begin{equation}
M_{JM}^{ij} = \left( \begin{array}{cc}
 0 & \hat{\mathcal{H}}^{(0)}_J-E_i \\
\hat{\mathcal{H}}^{(0)}_J+2\sigma_i^{(0)}\hat{K}_J\left[\sigma_i^{(0)}\right]-E_i & 0  
\end{array} \right)
\end{equation}
for $i=j$, and
\begin{equation}
M_{JM}^{ij} = \left( \begin{array}{cc}
 0 &0 \\
2\sigma_i^{(0)}\hat{K}_J\left[\sigma_j^{(0)}\right]  &  0
\end{array} \right)
\end{equation}
\end{subequations}
for $i\neq j$, where we have to replace $\sigma^{(0)2}$ with $\sum\limits_j \sigma^{(0)2}_j$ in the definition of $\hat{\mathcal{H}}^{(0)}_J$, see Eq.~(\ref{Eq.OpeH0_j}).
\end{widetext}

%%%%%%%%%%%%%%%%%%%%%%%%%%%%%%%
\section{Numerical Results}\label{Sec:numerical}
%%%%%%%%%%%%%%%%%%%%%%%%%%%%%%%
\begin{table*}
\caption{{\bf Stable configurations across different sectors of the theory:} Stability landscape as function of the self-interaction parameters $\lambda_n$ and $\lambda_s$. (g.s.): ground state, (s.b.): stability band, (n.e.): non existent. By {\it stability band} we mean  configurations that are only stable within certain ranges of the field amplitude. Configurations that do not exist for the single-field $s=0$ Gross-Pitaevskii-Poisson system are indicated in boldface.}
\begin{center}
\begin{tabular}{c c c c c c c c c c c }
\hline
\hline
& $\quad$ &\multicolumn{7}{c}{stationary} & $\quad$ & multi-frequency \\\cmidrule{3-9}
   & & \multicolumn{5}{c}{constant polarization} & $\quad$ & radial polarization & & \\
   \cmidrule{3-7}
   & & linear & & circular & & elliptical &  & & & 
\\ \hline \hline\\[-8pt]
$\lambda_n=0,\, \lambda_s = 0$ & & $n=0$ (g.s.) & $\quad$ & $n=0$ (g.s.)& $\quad$ & $n=0$ (g.s.)& & {\boldmath $n=0$} & & \textbf{(s.b.)} \\[1.5pt]
\hline 
\multirow{2}{*}{$\lambda_n> 0,\, \lambda_s = 0$} & & $n=0$ (g.s.)& $\quad$ & $n=0$ (g.s.)& $\quad$ & $n=0$ (g.s.)& & {\boldmath $n=0$} \textbf{(s.b.)} & & \multirow{2}{*}{\textbf{(s.b.)}} \\[1pt]
& & $n=1$ (s.b.)& $\quad$ & $n=1$ (s.b.) &  & $n=1$ (s.b.) & & {\boldmath $n=1$} \textbf{(s.b.)}  & & \\[2.5pt]
$\lambda_n < 0,\, \lambda_s = 0$ & & $n=0$ (s.b.)& $\quad$ & $n=0$ (s.b.)& $\quad$ & $n=0$ (s.b.)& & {\boldmath $n=0$} \textbf{(s.b.)} & & \textbf{(s.b.)} \\[1.5pt]
\hline 
\multirow{2}{*}{$\lambda_n=0,\, \lambda_s > 0$} & & \multirow{2}{*}{$n=0$ (g.s.)} & $\quad$ & $n=0$  (g.s.) & $\quad$ & \multirow{2}{*}{(n.e.)} & & \multirow{2}{*}{{\boldmath $n=0$} \textbf{(s.b.)}} & & \multirow{2}{*}{(n.e.)} \\
& & & $\quad$ & $n=1$ (s.b.) &  & & &  & & \\[2.pt]
$\lambda_n=0,\, \lambda_s < 0$ & & $n=0$ (s.b.) & $\quad$ & $n=0$ (s.b.)& $\quad$ & (n.e.) & & {\boldmath $n=0$} \textbf{(s.b.)} & & (n.e.) \\
\hline
\hline
\end{tabular}
\end{center}
\label{table.main}
\end{table*}

This section contains the main results of the paper, which consist of a classification of the spherical equilibrium configurations of the $s=1$ Gross-Pitaevskii-Poisson system according to their linear stability properties.

For concreteness, we proceed as follows: In Sec.~\ref{sec.free}, we explore the free theory, for which $\lambda_n =\lambda_s=0$, such that the self-interaction terms vanish. Next, in Sec.~\ref{sec.numberdensity}, we include the particle-particle self-interaction term ($\lambda_n\neq 0$ keeping $\lambda_s=0$), and in Sec.~\ref{sec.spindensity} the spin-spin self-interaction term ($\lambda_s\neq 0$ keeping $\lambda_n=0$). This allows us to isolate the two self-interaction terms and identify their effect on the equilibrium configurations. In both cases, we distinguish between two scenarios: an attractive particle-particle interaction ($\lambda_n>0$) and a repulsive ($\lambda_n<0$) one, and an antiferromagnetic spin-spin interaction ($\lambda_s>0$) and a ferromagnetic ($\lambda_s<0$) one, where we recall that a ground state exists only when $\lambda_n$ and $\lambda_s$ are positive, respectively. The general case in which the two self-interaction terms are present simultaneously ($\lambda_n,\lambda_s\neq 0$) is more involved and will not be addressed in this work.

To a large extend, our conclusions are derived from a numerical study of the linearized system~(\ref{eq.general system}) which we address using the same techniques as in Refs.~\cite{Nambo:2023yut,Nambo:2024gvs}. To make the presentation more accessible, we describe these methods in detail in App.~\ref{Sec:numerical.study} and dedicate this section to the presentation of the results. In all cases, our analysis includes perturbations with angular momentum numbers $J\le 5$, ensuring that the dominant low-multipole contributions have been considered.  We summarize our main findings in Table~\ref{table.main}.

%%%%%%%%%%%%%%%%%%%%%%%%%%%%%%%%%%%%%%%%%%
\subsection{Free theory}\label{sec.free}
%%%%%%%%%%%%%%%%%%%%%%%%%%%%%%%%%%%%%%%%%%
In absence of self-interactions ($\lambda^{phys}_n=\lambda^{phys}_s = 0$), the $s=1$ Gross-Pitaevskii-Poisson system~(\ref{eq.GPPs=1}) is invariant under the scaling transformation
\begin{equation}\label{eq.scaling.free}
t\to\lambda_*^{-1}t, \quad 
\vec{x}\to\lambda_*^{-1/2}\vec{x}, \quad 
\mathcal{U}\to\lambda_* \mathcal{U},\quad \vec{\psi} \to \lambda_*\vec{\psi},
\end{equation}
where $\lambda_*$ is an arbitrary and positive constant. This invariance can be traced back to the arbitrary scale $\lambda_*$ appearing in the variables~(\ref{eq.code.numbers1}), which disappears from the resulting dimensionless equations when $\lambda^{phys}_n$ and $\lambda^{phys}_s$ vanish. Since $\lambda_0 = 0$ in Eq.~(\ref{Eq:lambda0}), there exists a ground state for fixed $N$, which corresponds to a stationary and spherically symmetric solution of constant polarization and no nodes, $n=0$.

%%%%%%%%%%%%%%%%%%
\subsubsection{Stationary states of constant polarization}\label{sec.constant.free.results}
%%%%%%%%%%%%%%%%%%
\begin{figure*}
\centering
\includegraphics[width=18.cm]{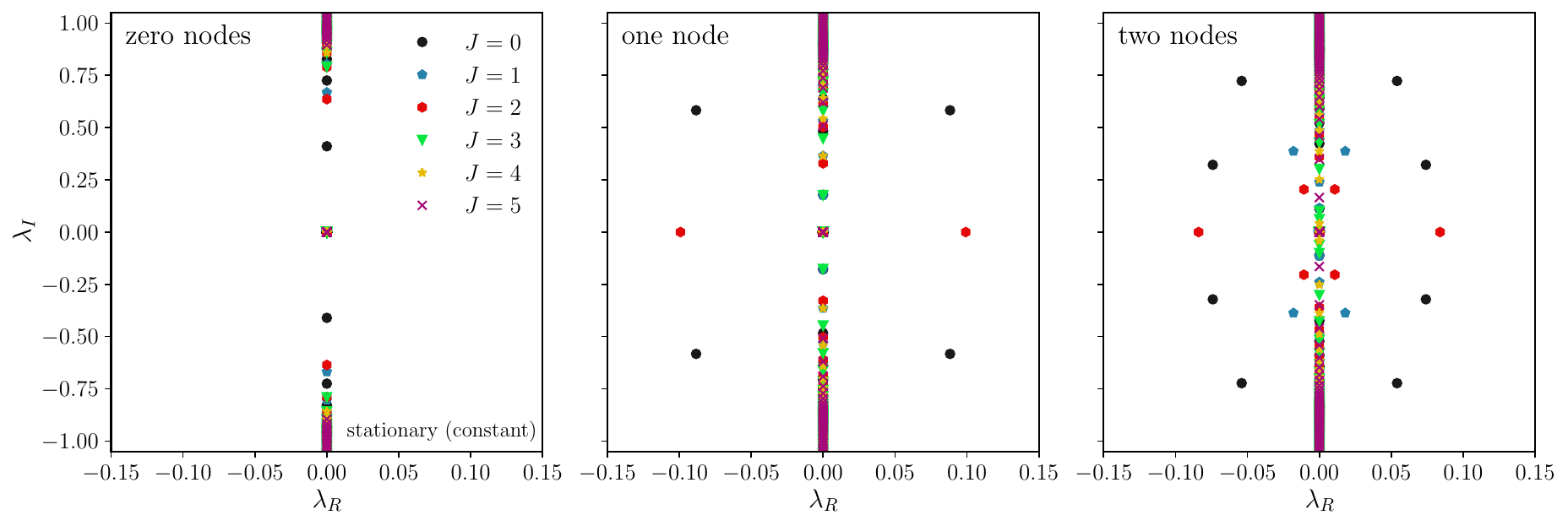}
\caption{
{\bf Eigenvalue spectra of stationary Proca stars with constant polarization (free theory):}
Spectra corresponding to the ground state ($n=0$, left panel) and two excited states ($n=1$, center; $n=2$, right) for configurations of constant polarization, unit amplitude $\sigma_0=1$, and perturbations up to $J\le 5$. Eigenvalues with $(\lambda_R, \lambda_I) \neq 0$ appear in quadruples ${\lambda, -\lambda, \lambda^*, -\lambda^*}$, while purely real and imaginary ones occur in pairs. Only the nodeless configuration ($n=0$) exhibits a purely imaginary spectrum and is therefore linearly stable. In contrast, excited states ($n\ge 1$) possess eigenvalues with nonzero real parts, implying linear instability.}\label{FigmodosSCPLinear}
\end{figure*}

In the free theory, constant polarization states are degenerate, and it is sufficient to study a single representative case. For definiteness, we will consider linearly polarized states.

To analyze the stability of linearly polarized Proca stars, we focus on the system~(\ref{eq.general system}), where $X_{JM}$, $Y_{JM}$, and $Z_{JM}$ are given by Eq.~(\ref{eq.defXYZ1}), and the matrices $M_{JM}^{ij}$ by Eqs.~(\ref{eq.matriz.linear}) with $\lambda_n=\lambda_s=0$. As shown in Sec.~\ref{sec.linear.pol}, these equations decouple into three independent subsystems. The first subsystem, $X_{JM}=M_{JM}^{11}X_{JM}$, coincides with that of Eq.~(35) in Ref.~\cite{Nambo:2024gvs} for nonrelativistic boson stars (once we set to zero the self-interaction constant in those equations, which corresponds to eliminate the terms with the $\pm$ signs in Eqs.~(36) of Ref.~\cite{Nambo:2024gvs}, whereas the other two subsystems, $Y_{JM}=M_{JM}^{22}Y_{JM}$ and $Z_{JM}=M_{JM}^{33}Z_{JM}$, are characterized in terms of self-adjoint operators $M_{JM}^{22}=M_{JM}^{33}$, which allows us to infer that all associated modes are purely oscillatory and no new instabilities appear, thereby recovering the results of Ref.~\cite{Nambo:2024gvs} for the $s=0$ Schr\"odinger-Poisson system.

Figure~\ref{FigmodosSCPLinear} presents the eigenvalue spectra of the system~(\ref{eq.general system}) for $n = 0$ (ground state), $n = 1$, and $n = 2$ linearly polarized Proca stars of unit  amplitude $\sigma_{0} = 1$. The analysis includes perturbations with angular momentum numbers $J = 0, 1, 2,3,4$, and $5$, and the results indicate that only the ground state does not exhibit unstable modes, given that for this configuration all eigenvalues are purely imaginary. In contrast, the configuration with one node presents $J=0$ and $J=2$ instabilities, while the configuration with two nodes is unstable for $J=0$, $J=1$, and $J=2$ perturbations. In both cases, no instabilities are observed for $J>2$.\footnote{In this paper we have studied perturbations with $J\le 5$, although in Ref.~\cite{Nambo:2024gvs} we extended the analysis to $J\le 12$ for boson stars, which share the same linear  stability behavior as constantly polarized Proca stars.\label{footnoteJ}} Note that the number of unstable $J=0$ modes is determined by the number of nodes of the background configuration, with each node introducing a pair of exponentially in time growing radial instabilities. (Remember that we count the number of instabilities by the number of modes with   positive real parts of $\lambda$.) This behavior is observed for all values of $n$ that we have explored.

Using the scaling symmetry~(\ref{eq.scaling.free}), we can extrapolate these results to configurations of arbitrary amplitude $\sigma_0$,  and conclude that, in the free theory, constantly polarized Proca stars with no nodes, which represent ground states for fixed $N$, are always stable, as shown analytically in App.~\ref{Sec:SecondVariation}, while those with nodes are unstable. This is consistent with the findings of Refs.~\cite{2002math.ph...8045H, Roque:2023sjl, Nambo:2023yut, Nambo:2024gvs} for nonrelativistic boson stars, including $\ell=0$ configurations. However, it is worth emphasizing that here the numerical solutions have been independently re-derived using a new code that solves the full system of Eqs.~(\ref{eq.general system}), rather than only the subsystem 
$X_{JM}=M_{JM}^{11}X_{JM}$ considered in our previous work~\cite{Nambo:2024gvs}. This also applies to other parts of the paper, where all the results were obtained using the same independent code that is adapted to handle the full spectrum of equilibrium configurations of a vector field.

%%%%%%%%%%%%%%%%%%%%%%%%
\subsubsection{Stationary states of radial polarization}\label{sec.free.circular}
%%%%%%%%%%%%%%%%%%%%%%%%
\begin{figure*}
\centering
\includegraphics[width=18.cm]{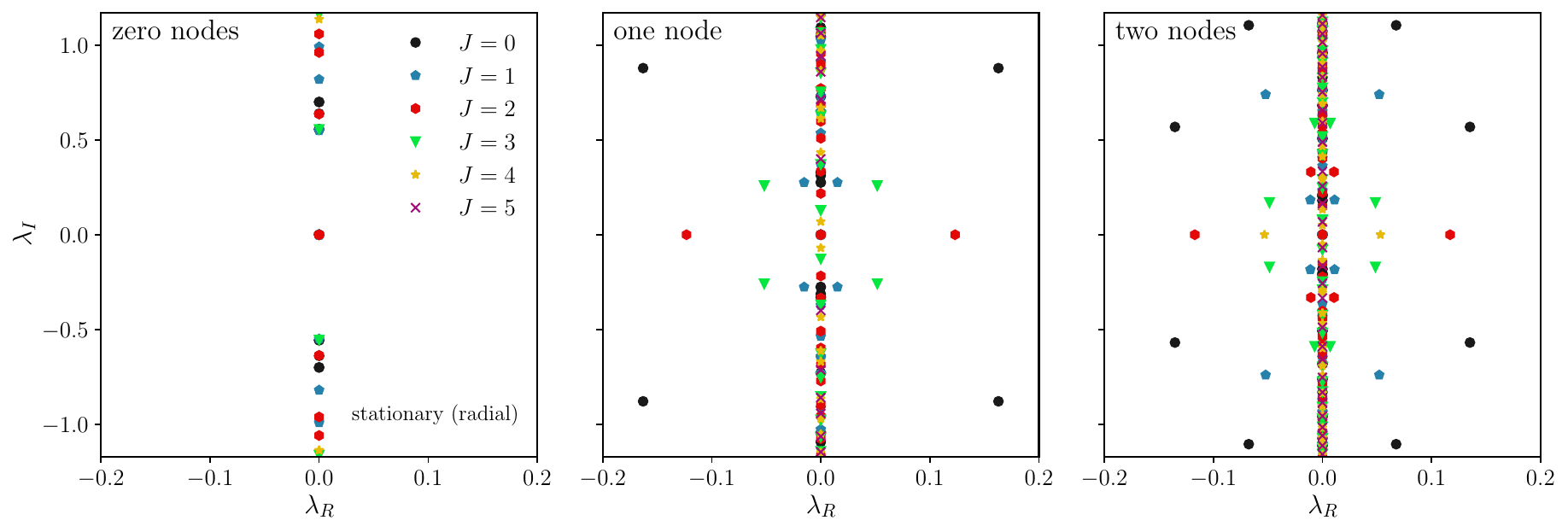}
\caption{{\bf Eigenvalue spectra of stationary Proca stars with radial polarization (free theory):} Same as in Fig.~\ref{FigmodosSCPLinear} but for radially polarized configurations. As for the constant polarization case, only the nodeless state is stable, although in this case it does not represent a ground state. }\label{FigmodosSCPRadial}
\end{figure*}

For fixed particle number $N$, and in the  absence of self-interaction terms, the energy of radially polarized Proca stars is always greater than that of the ground state; see, e.g., the central panel of Fig.~12 in Ref.~\cite{Nambo:2024hao}. This indicates that radially polarized configurations correspond to excited states of the $s = 1$ Schr\"odinger-Poisson system. Although a lower energy state exists, this does not necessarily imply that these configurations are unstable. As we shall see, some remain stable under linear perturbations, corresponding to local minima of the energy functional~(\ref{Eq.ConsEnergFunct}) for fixed $N$.

To analyze the linear stability of radially polarized Proca stars, we study the system~(\ref{eq.general system}), where $X_{JM}$, $Y_{JM}$, and $Z_{JM}$ are given by Eq.~(\ref{eq.defXYZ2}), and the matrices $M_{JM}^{ij}$ by Eqs.~(\ref{eq.matriz.radial}) with $\lambda_n=\lambda_s=0$. This system coincides with that of Eqs.~(41) in Ref.~\cite{Nambo:2023yut} for nonrelativistic $\ell=1$ boson stars, since in the nonrelativistic limit a  constant unitary transformation exists that relates radially polarized states to $\ell=1$ states, as shown in App.~D of Ref.~\cite{Nambo:2024hao}.

Figure~\ref{FigmodosSCPRadial} presents the eigenvalue spectra of the system~(\ref{eq.general system}) for $n = 0,\, 1$, and $2$ radial polarization states of unit amplitude $\sigma_{0} = 1$. As in the case with linear polarization, only the configuration with no nodes is free of unstable modes, while the configuration with one node  is unstable for $J=0, 1, 2$, and $3$, and the configuration with two nodes for $J=0,1,2,3$, and $4$. In both cases, no instabilities are observed for $J>4$.\footnote{In Sec.~IV.E of Ref.~\cite{Nambo:2023yut} it is proven that for high enough values of $J$, unstable modes are, in fact, excluded. In this reference we also excluded unstable modes with $3\le J\le 11$ for configurations with $n=0$ and $1$.} Again, the number of unstable $J=0$ modes appears to grow by two with each additional node of the wave function, a behavior that we have explicitly verified for the first few values of $n$. It is important to stress that, although the $n=0$ configuration is stable, it does not represent the ground state of the $s=1$ Gross-Pitaevskii-Poisson system for fixed $N$; rather it is expected to be the lowest energy stationary configuration for fixed $\hat{Q}=\frac{N}{3}\textrm{diag}(1,1,1)$ [see Eq.~(\ref{eq.globalQ})]. Again, we can extend these conclusions to arbitrary amplitudes $\sigma_0$ using the scaling~(\ref{eq.scaling.free}). This is consistent with the results reported in Refs.~\cite{Roque:2023sjl, Nambo:2023yut} for  $\ell=1$ boson stars.

%%%%%%%%%%%%%%%%%%%%%%%%
\subsubsection{Multi-frequency states}\label{sec.multi.free}
%%%%%%%%%%%%%%%%%%%%%%%%
\begin{figure}
\centering
\includegraphics[width=1.\linewidth]{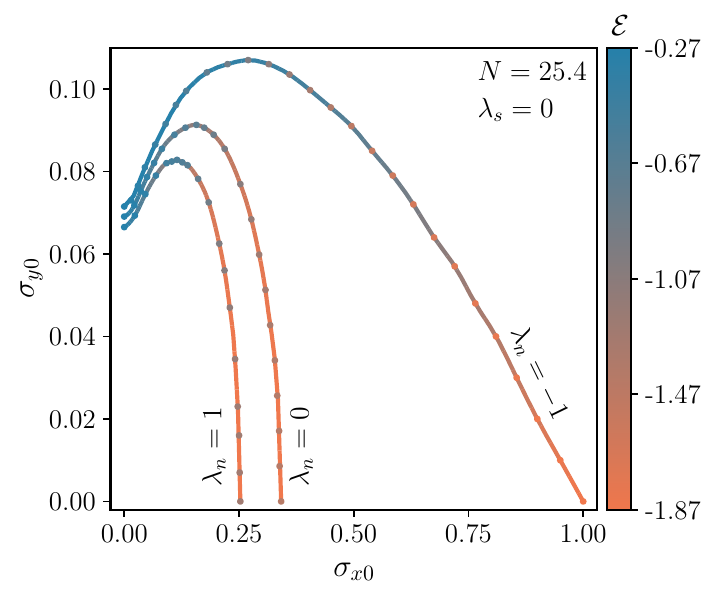}
\caption{{\bf Phase diagrams of multi-frequency states:} A family of two-component multi-frequency states with $(n_x,n_y)=(0,1)$ and $N=25.4$ in the attractive ($\lambda_n=-1$), free ($\lambda_n=0$), and repulsive ($\lambda_n=+1$) cases of the symmetry-enhanced sector of the theory ($\lambda_s=0$).}\label{FigProfMultifrequency}
\end{figure}

\begin{figure*}
\centering
\includegraphics[width=1.\linewidth]{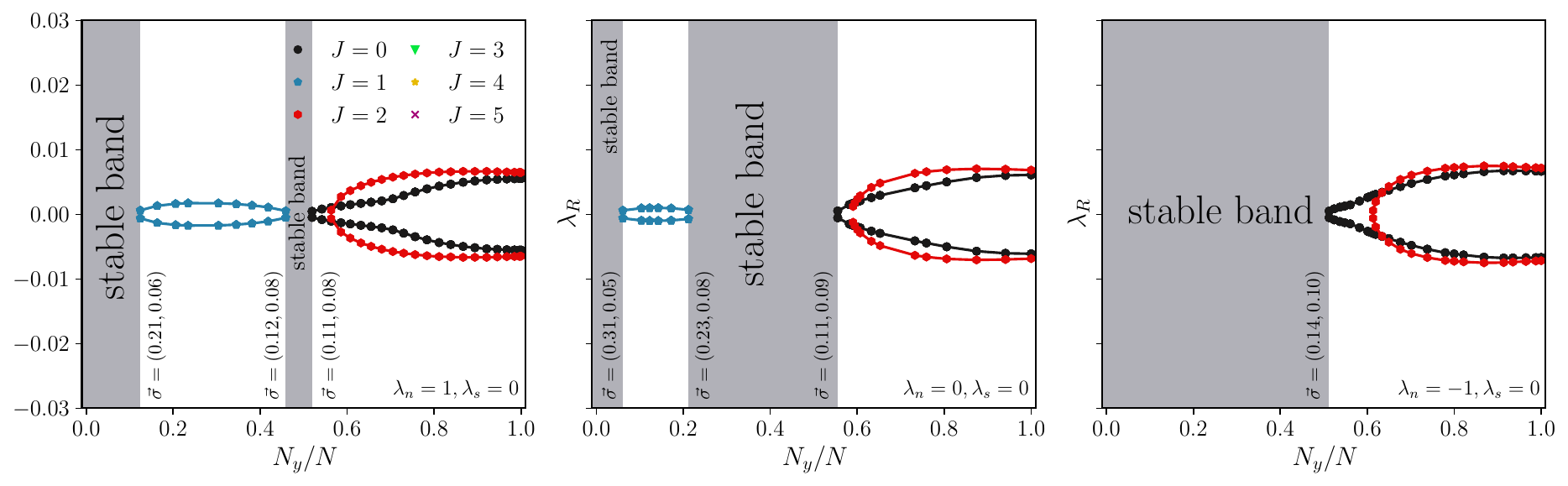}
\caption{{\bf Real eigenvalue spectra of two-component multi-frequency Proca stars:}
Spectra corresponding to the family presented in Fig.~\ref{FigProfMultifrequency}  (left: $\lambda_n=+1$, center: $\lambda_n=0$, right: $\lambda_n=-1$), for perturbations with $J\le 5$. The shaded regions indicate stability bands, where no unstable modes are observed. These configurations correspond to stable excited states.}\label{FigmodosMultifrequency}
\end{figure*}

The study of multi-frequency states is more involved than that of constant or radial polarization states, since in this case the three components of the field $\vec{\psi}$ can occupy different states, requiring  two sets of three numbers to fully specify the state of the system: $(n_x, n_y, n_z)$ and $(\sigma_{x0}, \sigma_{y0}, \sigma_{z0})$.\footnote{ Components with the same node numbers oscillate with the same frequency and can be removed by an internal rotation. This is the reason why we impose $n_x<n_y<n_z$ for multi-frequency configurations. A more detailed discussion can be found at the end of Sec.~IV.B of Ref.~\cite{Nambo:2024hao}.}

For definiteness, in this paper we focus on a family of configurations with node numbers $(n_x, n_y) = (0,1)$, for which the $z$-component vanishes and the central amplitudes $(\sigma_{x0},\sigma_{y0})$ yield a fixed particle number $N = 25.4$ (see the curve belonging to $\lambda_n = 0$ in Fig.~\ref{FigProfMultifrequency}).\footnote{The linear stability of multi-frequency states for which all three-components are non-zero will be analyzed in a follow-up paper.} It is important to note that, due to the scaling freedom~(\ref{eq.scaling.free}), our results generalize to any family of solutions with the same node numbers $(0,1)$, regardless of the value of $N$, because there is a one-to-one correspondence between states that preserve the same ratio among the components $\sigma_{i0}$ across different families. In particular, by choosing $\lambda_*=2.9$, we recover the family reported in Ref.~\cite{Nambo:2024hao}, with $N=43.5$ (see Figs.~$10$ and $15$ therein).

The  $N=25.4$ family of multi-frequency solutions begins with a nonzero $x$-component and vanishing $y$-component, $(\sigma_{x0},\sigma_{y0}) = (0.34, 0)$, which corresponds to a stable stationary state of constant polarization, that represents a ground state (see the left panel of Fig.~\ref{FigmodosSCPLinear}). Then, the particle number in the $y$-component is gradually increased while keeping the total particle number fixed at $N = 25.4$, leading to the final configuration in which all particles occupy the $y$-component: $(\sigma_{x0},\sigma_{y0})= (0, 0.069)$. This final state represents an unstable stationary configuration of constant polarization and higher energy than the ground state (see the center panel of Fig.~\ref{FigmodosSCPLinear}).

In this context, we expect the onset of instability along this family of solutions to occur when the number of particles in the $y$-component becomes sufficiently large to trigger the typical instabilities associated with the excited states of constant polarization. To identify this transition point, we define the new variable $N_y/N$, which quantifies the fraction of particles in the excited state $N_y$ relative to the total particle number $N$  and remains invariant under the scaling transformation~(\ref{eq.scaling.free}). The values $N_y/N = 0$ and $N_y/N = 1$ correspond to the extremal configurations $(\sigma_{x0},\sigma_{y0}) = (0.34, 0)$ and $(\sigma_{x0},\sigma_{y0}) = (0, 0.069)$, respectively.

The linear stability of multi-frequency states is determined by the system~(\ref{eq.general system}), where $X_{JM}$, $Y_{JM}$, and $Z_{JM}$ are given by Eq.~(\ref{eq.defXYZ1}), and the matrices $M_{JM}^{ij}$ by Eqs.~(\ref{Eq:MijJMMulti}) with $\lambda_n=\lambda_s=0$. The central panel of Fig.~\ref{FigmodosMultifrequency} shows the results of the eigenvalue problem~(\ref{eq.general system}) for different values of the angular momentum number $J$, with  particular attention on the unstable modes, identified by a nonvanishing real and positive part of the frequency $\lambda$. As expected, no unstable modes are found for the extremal configuration $N_y/N = 0$, while at the other extreme of the parameter space ($N_y/N = 1$) two unstable modes are observed for $J = 0$ and $J = 2$. These results are consistent with our findings of  Fig.~\ref{FigmodosSCPLinear}. Interestingly, radial instabilities ($J = 0$) begin to appear when the particle number becomes approximately evenly distributed between the $x$ and $y$ components, i.e. for $N_y/N \gtrsim 0.55$. This behavior is consistent with that reported in Ref.~\cite{Urena-Lopez:2010zva} for mixed Newtonian boson stars (which are described by the same system of equations as multi-frequency configurations),\footnote{Note that the expression for the radial part of the Laplace operator in Eqs.~(20) and~(22) of Ref.~\cite{Urena-Lopez:2010zva} contains a typo. However, we have verified that the numerical results concerning the equilibrium mixed states reported in their Table~I coincide with the ones  obtained in this paper.} where radial stability was studied using numerical evolutions. However, in addition to the purely radial perturbations, we find $J=1$ instabilities within the radially stable region for $0.06 \lesssim  N_y/N \lesssim  0.21$, and $J=2$ instabilities for $N_y/N \gtrsim 0.55$, within the region that already contains radially unstable modes. No unstable modes with $J>2$ were identified. This is consistent with what we have observed for other configurations, where it is shown that for sufficiently large values of $J$, unstable modes are absent.

%%%%%%%%%%%%%%%%%%%%%%%%%%%%%%%%%%
\subsection{Self-interactions proportional to the squared number density}\label{sec.numberdensity}
%%%%%%%%%%%%%%%%%%%%%%%%%%%%%%%%%%
Self-interaction terms break the invariance of the $s=1$ Gross-Pitaevskii-Poisson system under the scaling  transformation~(\ref{eq.scaling.free}). This complicates the analysis in two ways. On the one hand, the perturbation equations must be solved for the specific values of $\lambda_n^{phys}$ and $\lambda_s^{phys}$ of interest, and in addition, the procedure must be repeated for each value of the amplitude $\sigma_0$. Taken together, these requirements make a systematic exploration of the parameter space considerably more demanding than in the free theory.

In this section, we concentrate on the case in which $\lambda_n^{phys}\neq 0$ and $\lambda_s^{phys}=0$. To proceed, we use the freedom in Eqs.~(\ref{eq.code.numbers1}) to set $\lambda_* = |\lambda^{phys}_n|$. This fixes the dimensionless self-interaction  constants to $\lambda_n = \pm 1$ and $\lambda_s = 0$, where the $\pm$ sign in $\lambda_n$ distinguishes between the repulsive $(+)$ and attractive $(-)$ cases, and reduces considerably the number of configurations to be studied independently. When the self-interaction term is repulsive, for fixed $N$, there is a ground state configuration which corresponds to a stationary and spherically symmetric solution of constant polarization and no nodes. However, when the self-interaction is attractive, a lowest energy state does not exist.  

%%%%%%%%%%%%%%%%%
\subsubsection{Stationary states of constant polarization}\label{sec.lambdan.constant}
%%%%%%%%%%%%%%%%%
\begin{figure*}
\centering
\includegraphics[width=1.\linewidth]{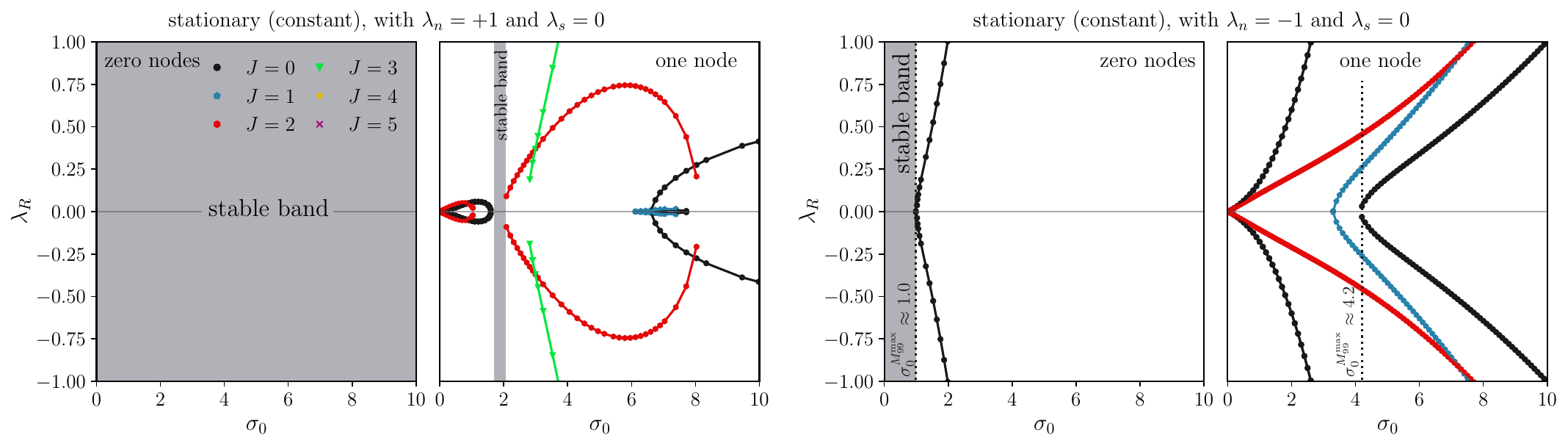}
\caption{{\bf Real eigenvalue spectra  of stationary Proca stars with constant polarization ($\lambda_n\neq0$, $\lambda_s=0$):} Spectra corresponding to self-interacting stationary states of constant polarization ($\lambda_n = +1$, left panel; $\lambda_n=-1$, right panel), for perturbations with $J\le 5$.  The shaded regions indicate stability bands where no unstable modes are observed. Vertical dotted lines correspond to maximum mass configurations.}\label{Fig.StatConstPolarLambdan}
\end{figure*}

In the absence of spin-spin self-interactions,  constant polarization states remain degenerate and we can restrict ourselves to the case of linear polarization. 

The stability of these states is determined by the system~(\ref{eq.general system}), where $X_{JM}$, $Y_{JM}$, and $Z_{JM}$ are given by Eq.~(\ref{eq.defXYZ1}), and the matrices $M_{JM}^{ij}$ by Eqs.~(\ref{eq.matriz.linear}) with $\lambda_n=\pm 1$ and $\lambda_s=0$. As for the free theory, these equations decouple into three independent subsystems: the first  one, $X_{JM}=M_{JM}^{11}X_{JM}$, coincides with that of Eqs.~(35) in Ref.~\cite{Nambo:2024gvs}, which describes nonrelativistic self-interacting  boson stars, while the other two, $Y_{JM}=M_{JM}^{22}Y_{JM}$ and $Z_{JM}=M_{JM}^{33}Z_{JM}$, involve self-adjoint operators $M_{JM}^{22}$ and $M_{JM}^{33}$ and do not introduce new unstable modes. 

To proceed, we must analyze the repulsive ($\lambda_n=+1$, $\lambda_s=0$) and attractive ($\lambda_n=-1$, $\lambda_s=0$) cases separately. Moreover, since  the scaling symmetry~(\ref{eq.scaling.free}) is lost,  each amplitude $\sigma_0$ must be studied individually. Figure~\ref{Fig.StatConstPolarLambdan} presents the eigenvalue spectra of the system~(\ref{eq.general system}) for $n=0$  and $n=1$ linearly polarized Proca stars as functions of $\sigma_0$ in the repulsive (left panel) and the attractive (right panel) cases. Only eigenvalues  with nonzero real part $\lambda_R$ are shown, as they are the responsible of the unstable modes, and the analysis is restricted to perturbations with $J\le 5$ (see, however, footnote~\ref{footnoteJ}).

Our results indicate that, for repulsive self-interactions, nodeless configurations (which correspond to ground states for fixed $N$) do not exhibit unstable modes. This is consistent with the idea that ground state configurations are always stable, and agrees with the results obtained for the free theory and App.~\ref{Sec:SecondVariation}. The situation becomes more interesting for configurations with one node. As shown in Fig.~\ref{Fig.StatConstPolarLambdan}, when the amplitude of the field is small enough, we recover the same $J=0$ and $J=2$ instabilities that we identified in Sec.~\ref{sec.constant.free.results} for the free theory. This behavior arises because the self-interaction terms involve higher powers of the field, making their contribution negligible at small amplitudes. However, as $\sigma_0$ increases, the eigenvalue spectra of the system~(\ref{eq.general system}) deviate from that of the free theory. In particular, new unstable modes with $J=1$ and $J=3$ appear for certain values of the amplitude $\sigma_0$, but notably, a region emerges where  no unstable modes are present, giving rise to a stability band. For higher values of $n$, we do not find additional stability bands, and these states remain unstable, as it also happens in the free theory. This agrees with what we found in Ref.~\cite{Nambo:2024gvs} for nonrelativistic self-interacting boson stars, and is also consistent with the results reported in Refs.~\cite{Sanchis-Gual:2021phr,Brito:2023fwr,Brito:2025rld}.

In contrast, for attractive self-interactions, stability is restricted to states with no nodes and  small field amplitude, $\sigma_0\lesssim 1$. For $n=0$ and amplitudes larger than one, the configurations develop two $J=0$ unstable modes, which are released at the maximum of the $M_{99}$ vs. $R_{99}$ curve, indicated by a vertical doted line in the right panel of Fig.~\ref{Fig.StatConstPolarLambdan}. In this case, the energy is unbounded from below, and these states do not correspond to the ground state of the system, which, in fact, does not exist for an attractive particle-particle self-interaction term. For higher values of $n$ all linear polarization states  are found to be unstable.

These results indicate that a particle-particle self-interaction term affects linearly polarized states in two ways: For repulsive self-interactions, it induces a stability band in $n=1$ configurations, and for attractive self-interactions, it destabilizes large amplitude $n=0$ configurations. The stability of the remaining states is unchanged from the free theory.

%%%%%%%%%%%%%%%%%%%%%%%%%%%%%%%%%%%%
\subsubsection{Stationary states of radial polarization}
%%%%%%%%%%%%%%%%%%%%%%%%%%%%%%%%%%%%
\begin{figure*}
\centering
\includegraphics[width=1.\linewidth]{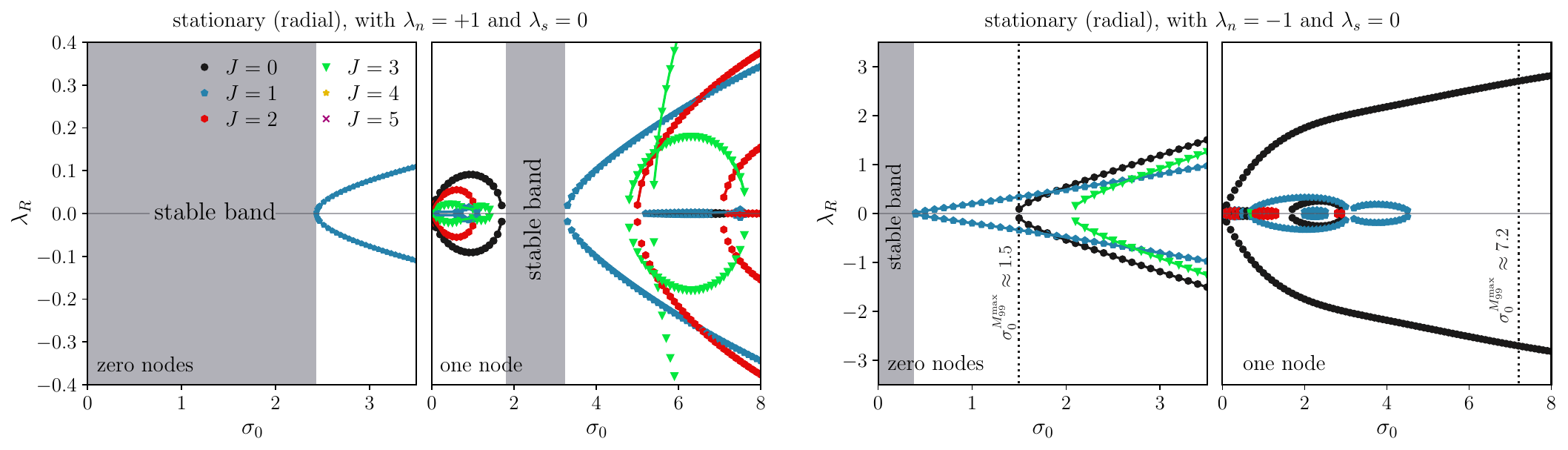}
\caption{{\bf Real eigenvalue spectra  of stationary Proca stars with radial polarization  ($\lambda_n\neq0$, $\lambda_s=0$):} Same as in Fig.~\ref{Fig.StatConstPolarLambdan} but for stationary states of radial polarization. Note that instabilities of nodeless states (first and third panels) are not triggered at a maximum mass configuration.}\label{Fig.StatRadialPolarLambdan}
\end{figure*}

If the spin–spin self-interaction term vanishes, radially polarized states with $\lambda_n>0$ remain more energetic than the ground state (see, e.g., the left panel of Fig.~12 in Ref.~\cite{Nambo:2024hao}), and can thus be identified as excited states. This is the same behavior that we found in the free theory. However, when $\lambda_n<0$, the situation is more  complicated. On the one hand, in this case the energy is unbounded from below and there does not exist a ground state. Moreover, nodeless constant polarization states do not always correspond to the lowest energy equilibrium configuration for fixed $N$ (see, e.g., the right panel of Fig.~12 in Ref.~\cite{Nambo:2024hao}), and in some instances nodeless radial polarization states are in fact  less energetic.

To analyze the linear stability of radially polarized Proca stars, we consider the system~(\ref{eq.general system}), where $X_{JM}$, $Y_{JM}$, and $Z_{JM}$ are given by Eq.~(\ref{eq.defXYZ2}), and the matrices $M_{JM}^{ij}$ by Eqs.~(\ref{eq.matriz.radial}) with $\lambda_n=\pm 1$ and $\lambda_s=0$. To proceed, we follow the same strategy as in Sec.~\ref{sec.lambdan.constant}. Figure~\ref{Fig.StatRadialPolarLambdan} presents the eigenvalue spectra of the system~(\ref{eq.general system}) for $n=0$ and $n=1$ radially polarized Proca stars as functions of $\sigma_0$ in the repulsive (left panel) and the attractive (right panel) cases.

Our results show that, for repulsive self-interactions, configurations with no nodes are stable only for amplitudes $\sigma_0 \lesssim 2.44$, developing a  $J=1$ instability for larger values of $\sigma_0$. This behavior differs from that of the constant polarization case, in which nodeless configurations are stable for any amplitude $\sigma_0$. This is because, as mentioned previously, radially polarized states are not ground states of the $s=1$ Gross-Pitaevskii-Poisson system, so there is no {\it a priori} expectation that they should be stable. As in other cases, the effect of the self-interactions is negligible at small amplitudes, where we recover the stable states that we identified for the free theory. However, as the amplitude increases, the self-interaction triggers a $J=1$ instability, which makes these configurations unstable. 

For configurations with one node, a similar behavior is observed, although with different consequences. In this case, at small amplitudes, we reproduce the unstable $J=0,\,1,\,2$ and $3$ modes that we identified in Sec.~\ref{sec.free.circular}. However, as the amplitude increases and the self-interaction effects become more significant, the unstable modes disappear for certain values of the amplitude, giving rise to a stability band in the range $1.7\lesssim\sigma_0\lesssim 3.3$. This band is analogous to that found in the constant polarization case, but is broader here.

For attractive self-interactions,  only states with \mbox{$\sigma_0\lesssim 0.4$} and no nodes are stable. Contrary to the stationary states of constant polarization, here  the instability is not triggered at the maximum of the $M_{99}$ vs. $R_{99}$ curve, marked by the vertical dotted line in Fig.~\ref{Fig.StatRadialPolarLambdan}, which signals the onset of the $J=0$ instability.
Instead, at lower amplitudes, a non-radial $J=1$ unstable mode is already present and is primarily responsible for the emergence of the instability. For higher values of $n$, all radial polarization states  are found to be unstable. This is similar to what happens for constant polarization states.

In summary, a particle-particle self-interaction term induces a $J=1$ instability in nodeless radially polarized states, restricting stability to specific  bands. In particular, for repulsive self-interactions, stable bands appear in $n=0$ and $n=1$ states, while for attractive self-interactions only $n=0$ states exhibit a stable region.

%%%%%%%%%%%%%%%%%%%%%%%%%%%%%%%%%%%%%%%%
\subsubsection{Multi-frequency states}
%%%%%%%%%%%%%%%%%%%%%%%%%%%%%%%%%%%%%%%%
Since $\lambda_s=0$, multi-frequency states are still allowed when $\lambda_n = \pm 1$. We now examine how the particle-particle self-interaction term affects the $N=25.4$ family of multi-frequency states introduced in Sec.~\ref{sec.multi.free} (see Fig.~\ref{FigProfMultifrequency}). Due to the loss of scaling freedom, each value of $N$ must be necessarily studied independently. However, we expect that the qualitative behavior observed here remains valid for other families of solutions with different particle number.

The linear stability of multi-frequency states is determined by the system~(\ref{eq.general system}), where $X_{JM}$, $Y_{JM}$, and $Z_{JM}$ are given by Eq.~(\ref{eq.defXYZ1}), and the matrices $M_{JM}^{ij}$ by Eqs.~(\ref{Eq:MijJMMulti}) with $\lambda_n=\pm1$ and $\lambda_s=0$.
The left and right panels of Fig.~\ref{FigmodosMultifrequency} show the results of  this eigenvalue problem for  repulsive ($\lambda_n=+1$) and attractive ($\lambda_n=-1$) interactions, respectively. In both cases stability bands are observed for certain ranges of $N_y/N$. Radial $J=0$ unstable modes appear around $N_y/N \approx 0.55$, while no unstable modes with $J>2$ were found. Remarkably, the $J=2$ modes are the last to appear, no matter the sign of the self-interaction. In the repulsive case, unstable $J=1$ modes split the radially stable band into two narrow bands, as in the free theory (central panel), with the notable difference that in the self-interacting case the unstable $J=1$ band is larger and its eigenvalues are higher. In contract, it appears that an attractive particle-particle self-interaction term removes the $J=1$ unstable modes and  closes this instability band. 

%%%%%%%%%%%%%%%%%%%%%%%%%%%%%%%%%%%%%%%%%%%%%%%%%%%%
\subsection{Self-interactions proportional to the squared spin density}\label{sec.spindensity}
%%%%%%%%%%%%%%%%%%%%%%%%%%%%%%%%%%%%%%%%%%%%%%%%%%%%
We now turn to the case where $\lambda^{phys}_n = 0$ and $\lambda_s^{phys}\neq0$. Without loss of generality, we can set $\lambda_* = |\lambda_s^{phys}|$ in Eqs.~(\ref{eq.code.numbers1}), which fixes the dimensionless self-interaction constants to  $\lambda_n = 0$ and $\lambda_s = \pm 1$, where the plus and minus signs correspond to the antiferromagnetic and ferromagnetic cases, respectively. The spin-spin self-interation term  eliminates the possibility of multi-frequency states, which simplifies the analysis. However, it also breaks the degeneracy among constant polarization states, and linear and circular polarizations must be treated separately. When the self-interaction term is antiferromagnetic ($\lambda_s=+1$), the system admits a ground state configuration for fixed $N$, corresponding to a stationary spherically symmetric solution of constant linear polarization and no nodes, $n=0$. In contrast, for a ferromagnetic self-interaction ($\lambda_s=-1$) the energy is unbounded from below and no lowest energy state exists.

The stability of Proca stars is determined by Eqs.~(\ref{eq.general system}), which depend partly on the underlying background configurations. These are obtained from Eqs.~(\ref{s=1GPP.stationary}), where the self-interactions are codified in the constant $\lambda_n+\alpha\lambda_s$. For $\lambda_n=0$, linear ($\alpha=\gamma=0$) and radial ($\alpha=0$, $\gamma=1$) polarization states reduce to their corresponding free theory solutions, see Secs.~\ref{sec.constant.free.results} and~\ref{sec.free.circular}, respectively. Meanwhile, circular ($\alpha=1$, $\gamma=0$) polarization states reduce to those of the self-interacting theory with $\lambda_n\neq 0$ and $\lambda_s=0$ described in Sec.~\ref{sec.lambdan.constant}, upon exchanging $\lambda_s$ with $\lambda_n$. However, this  correspondence does not, in general, extend beyond the equilibrium solutions, since the self-interaction terms also enter the perturbation equations in combinations different from $\lambda_n+\alpha\lambda_s$ (see the matrices $M_{JM}^{ij}$ defined in~(\ref{eq.matriz.linear}),~(\ref{eq.matriz.circular}), and~(\ref{eq.matriz.radial})). This implies that the stability of the sector $\lambda_n=0$, $\lambda_s= \pm1$ cannot be directly inferred from the previous cases and therefore
requires an independent analysis which we present in this section. 

%%%%%%%%%%%%%%%%%%%%%%%%%%%%%%%%%%%%%%%%%%%%%%%%
\subsubsection{Stationary states of constant linear polarization}
%%%%%%%%%%%%%%%%%%%%%%%%%%%%%%%%%%%%%%%%%%
The spin-spin self-interaction term removes the degeneracy among constant polarization states and leaves the linear and circular configurations as the only viable ones. In this section, we analyze linearly polarized Proca stars, leaving the circular case to Sec.~\ref{sec.circulas.lambdas}.

The stability of stationary states of constant linear polarization is determined by the system~(\ref{eq.general system}), where $X_{JM}$, $Y_{JM}$, and $Z_{JM}$ are given by Eq.~(\ref{eq.defXYZ1}), and the matrices $M_{JM}^{ij}$ by Eqs.~(\ref{eq.matriz.linear}) with $\lambda_n= 0$ and $\lambda_s=\pm 1$. As in previous cases, these equations decouple into three independent subsystems. The first subsystem, $X_{JM}=M_{JM}^{11}X_{JM}$, is identical to the same subsystem of Sec.~\ref{sec.constant.free.results}, and therefore exhibits the same spectrum of solutions, including the unstable modes. The other two subsystems, $Y_{JM}=M_{JM}^{22}Y_{JM}$ and $Z_{JM}=M_{JM}^{33}Z_{JM}$, however, differ from those of Sec.~\ref{sec.constant.free.results} due to the term proportional to $\lambda_s$ in Eq.~(\ref{eq.matriz.linear2}), which was  not previously present. This additional term renders the matrices $M_{JM}^{22}$ and $M_{JM}^{33}$ non-self-adjoint, and in this case we cannot guarantee that they do not introduce new unstable modes.

This may modify the behavior described in Sec.~\ref{sec.constant.free.results}, and for that reason we have solved the full system that includes the new term proportional to $\lambda_s$. However, for the range of amplitudes that we have explored, $0<\sigma_0<80$, the unstable modes of the new system coincide with those  found in Sec.~\ref{sec.constant.free.results}, and therefore also with those of the first subsystem. In this case we can use~(\ref{eq.scaling.free}) to scale the solutions to arbitrary values of $\sigma_0$, at least for those amplitudes that match the free theory, and for that reason we have not included a new figure for the real eigenvalue spectra. This shows that, for these amplitudes, the second and third subsystems, despite involving non-self-adjoint operators, do not generate additional instabilities. This suggests that the spin-spin self-interaction term has no significant impact on the stability of stationary states of constant linear polarization.

%%%%%%%%%%%%%
\subsubsection{Stationary states of constant circular polarization}\label{sec.circulas.lambdas}
%%%%%%%%%%%%%%%%
\begin{figure*}
\centering
\includegraphics[width=1.\linewidth]{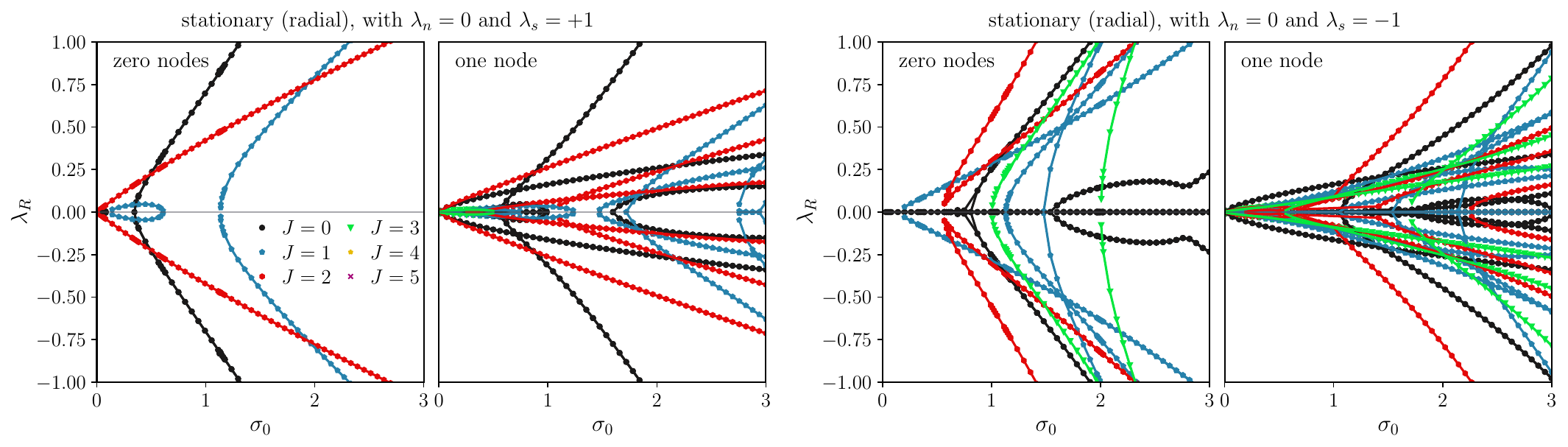}
\caption{{\bf Real eigenvalue spectra of stationary Proca stars with radial polarization ($\lambda_n=0$, $\lambda_s\neq 0$):} Same as in Fig.~\ref{Fig.StatRadialPolarLambdan} but for spin-spin self-interaction. Note that for small field amplitudes, $\sigma_0\lesssim0.01$, the configurations without nodes are stable in the antiferromagnetic and ferromagnetic cases, although this is barely visible from the figures.}\label{Fig.StatRadPolarLambdas}
\end{figure*}

The stability of stationary states of constant circular polarization is determined by the system~(\ref{eq.general system}), where $X_{JM}$, $Y_{JM}$, and $Z_{JM}$ are given by Eq.~(\ref{eq.defXYZ1}), and the matrices $M_{JM}^{ij}$ by Eqs.~(\ref{eq.matriz.circular}) with $\lambda_n= 0$ and $\lambda_s=\pm 1$. In this case, Eqs.~(\ref{eq.general system}) decouple into two subsystems, one for the variables $X_{JM}$ and $Y_{JM}$, and another for $Z_{JM}$. While no direct correspondence is evident between the equations that describe the perturbations for $\lambda_n=\pm 1$, $\lambda_s=0$ and  $\lambda_n= 0$, $\lambda_s=\pm 1$, for the range of amplitudes that we have explored, $0<\sigma_0<80$, we recover the same spectrum of unstable modes as in Sec.~\ref{sec.lambdan.constant}, see Fig.~\ref{Fig.StatConstPolarLambdan}. This leads us to conclude that, for circularly polarized states, the effect of $\lambda_s$ is equivalent to that of $\lambda_n$, although we have not been able to establish this property analytically.

%%%%%%%%%%%%%%%%%%%%%%%%%%%%%%%%%%%%%%%%%%%%
\subsubsection{Stationary states of radial polarization}
%%%%%%%%%%%%%%%%%%%%%%%%%%%%%%%%%
The stability of stationary states of radial polarization is determined by the system~(\ref{eq.general system}), where $X_{JM}$, $Y_{JM}$, and $Z_{JM}$ are given by Eq.~(\ref{eq.defXYZ2}), and the matrices $M_{JM}^{ij}$ by Eqs.~(\ref{eq.matriz.radial}) with $\lambda_n= 0$ and $\lambda_s=\pm 1$. These equations differ from those of the free theory only in a term proportional to $\lambda_s$ that appears in the upper right element of the matrix $M_{JM}^{22}$. However, in this case, this minor modification produces a markedly different behavior compared to the free theory.

Figure~\ref{Fig.StatRadPolarLambdas} shows the real eigenvalue spectra of the system~(\ref{eq.general system}) for $n=0$ and $n=1$ radially polarized Proca stars as functions of $\sigma_0$ in the antiferromagnetic (left panel) and ferromagnetic (right panel) cases. As for particle-particle self-interactions, spin-spin self-interactions  destabilize some states with radial polarization, although the effect is even more pronounced now. In fact, stability is restricted to a very narrow subset of nodeless states with amplitudes $\sigma_0 \lesssim 0.01$, outside of which all the studied states have been found to be unstable. This threshold is so small that the corresponding stability band is not visible in Fig.~\ref{Fig.StatRadPolarLambdas}. For $n>0$, no stable states have been identified. Our results reveal the unstable modes of the free theory, along with additional ones originating from the fact that the matrices $M_{JM}^{22}$ and $M_{JM}^{33}$ are not self-adjoint as a consequence of a nonvanishing $\lambda_s$. Contrary to what happens for linear and circular polarization states, the the radial case is much more sensitive to the presence of $\lambda_s$.

In conclusion, while for linear and circular polarizations a nonvanishing $\lambda_s$ has little  impact on the stability of the solutions, the situation is markedly different for the radial polarization. In this case, the system is far more sensitive to the spin-spin self-interaction, which destabilizes configurations that would otherwise remain stable, such as those associated with the free theory and with particle-particle self-interactions.

%%%%%%%%%%%%%%%%%%%%%%%%%%%%%%%
\section{Conclusions}\label{Sec:Conclussions}
%%%%%%%%%%%%%%%%%%%%%%%%%%%%%%%
In this paper we have studied the stability of the spherical equilibrium configurations of the $s=1$ Gross-Pitaevskii-Poisson system
under generic linear perturbations. These solutions consist of stationary states of constant or radial polarization, as well as  multi-frequency states when $\lambda_s=0$. The main results, which provide a classification of the stable states as function of the self-interaction constants $\lambda_n$ and $\lambda_s$, are summarized in Table~\ref{table.main}.

For $\lambda_0\ge 0$ [see Eq.~(\ref{Eq:lambda0})], the energy of the system is bounded from below and there exists a ground state configuration which  possesses the lowest energy for fixed particle number $N$.  This state is stationary and spherically symmetric, with a monotonically decreasing profile of its wave function which has constant polarization, that can be linear (the antiferromagnetic phase) or circular (the ferromagnetic phase), depending on the sign of $\lambda_s$ (in the symmetry enhanced sector $\lambda_s=0$ the two phases are degenerate). The ground state is always stable against small perturbations, and it generates a gravitational field that is indistinguishable from the one associated with the ground state of a boson star.

Therefore, any difference between the gravitational field of Proca stars and boson stars must originate from their excited states. While such states are generally expected to be unstable (see, however, Refs.~\cite{Sanchis-Gual:2021phr,Brito:2023fwr,Brito:2025rld,Nambo:2024gvs} for exceptions in  self-interacting single-field scalar theories), in the case of Proca stars stability is not limited to the ground state, and certain excited configurations remain stable, even in the free theory.

In the absence of self-interactions, nodeless radially polarized Proca stars and certain multi-frequency states, which are more energetic than the ground state, are stable. These configurations have no analog in classical single-scalar-field theories, although they correspond, respectively, to $\ell=1$ boson stars~\cite{Roque:2023sjl,Nambo:2023yut} and to multi-state boson stars~\cite{Urena-Lopez:2010zva} in theories with more than one scalar field, or to configurations of a single quantum scalar field where particles occupy different energy levels~\cite{Alcubierre:2022rgp}.

Self-interaction terms affect the stable and unstable configurations of the free theory in different ways. In particular, we have studied the cases where the particle-particle self-interaction term $\lambda_n$ and the spin-spin self-interaction term $\lambda_s$ are considered separately. On the one hand, we have found that they do not affect the stability of the ground state, which remains stable in all cases as long as $\lambda_0$ is positive. Regarding the excited states, we have identified that a repulsive particle-particle self-interaction term ($\lambda_n>0$) induces a stability band in the $n=1$ stationary states of constant and radial polarization, destabilizes  large-amplitude $n=0$ radial states, and enhances the instability of multi-frequency states, although some stable configurations still persist. An antiferromagnetic spin-spin self-interaction term ($\lambda_s>0$), on the other hand, induces a stability band in the $n=1$ circularly polarized states and transforms nearly all radially polarized states into unstable ones.

When $\lambda_0<0$, the energy of a self-gravitating vector field is unbounded from below and no ground state configuration exists. This scenario should be regarded as a simplified limit of a more fundamental theory, in which higher-order self-interaction terms would ensure that the energy is bounded. The extended theory would then support equilibrium solutions, such as the ground state, that cannot be captured by our  effective model. However, this simplified model offers an accurate description of configurations with small field amplitudes, for which the contribution of the higher-order operators becomes negligible. In this limit, we have found that certain configurations, which  correspond to excited states of the complete theory, remain stable.

Although a stability analysis of the equilibrium configurations considered in this article can, in principle, also be studied by means of numerical evolutions, the linear analysis presented in this work offers a significant benefit given that it identifies in advance the regions of parameter space where instabilities arise. This makes the search for unstable configurations considerably more systematic and efficient, especially in cases that require very long evolution times. The results of nonlinear numerical simulations will be presented in a companion paper.

The stability of relativistic Proca stars was recently studied in Ref.~\cite{Herdeiro:2023wqf}, where it was found that the lowest energy configuration of the Einstein-Proca theory develops an axisymmetric, prolate geometry and that spherically symmetric, radially polarized states are unstable. Although this might seem to be incompatible with our results, the two analyses probe complementary physical regimes.
In the fully relativistic theory, the anisotropic stresses of the vector field can significantly distort the spacetime geometry and drive the ground state to a non-spherical configuration. In contrast, in the nonrelativistic limit that we have explored in this paper, the gravitational field is weak and the metric perturbations responsible for such shape deformations are thus expected to be strongly suppressed. Consequently,  the ground state configurations of Ref.~\cite{Herdeiro:2023wqf} are presumed to approach the spherically
symmetric constant polarized states found in this work when relativistic corrections are turned off.
A similar reasoning applies to the instability of radially polarized states, where the mechanism that destabilizes these configurations is likely tied to relativistic degrees of freedom that are absent (or parametrically suppressed) in the nonrelativistic theory. These observations make clear that a detailed study of the transition between the relativistic and nonrelativistic regimes should be interesting. We hope to address this question in future work.

%%%%%%%%%%%
\begin{acknowledgments}
%%%%%%%%%%%
We thank Jos\'e Luis Medina Garc\'ia for his help in validating the results of this paper with an evolution code. 
E.C.N., G.D.A. and E.P.G. were supported by SECIHTI predoctoral scholarships, and A.A.R. by the postdoctoral fellowship ``Estancias Posdoctorales por México para la Formación y Consolidación de las y los Investigadores por México''. A.D.T., A.A.R. and O.S. acknowledges support from SECIHTI-SNII. Additional support was provided to A.D.T. by the DAIP project CIIC 2024 198/2024, and to O.S. by CIC grant No.~18315 of the Universidad Michoacana de San Nicolás de Hidalgo. We also acknowledge the use of the computing server COUGHS from the UGDataLab at the Physics Department of Guanajuato University.
\end{acknowledgments}

%%%%%%%%%%%
\appendix
%%%%%%%%%%%

%%%%%%%%%%%%%%%%%%%%%%%%%%%%%%%
\section{Variations of the energy functional and stability of ground states}\label{Sec:SecondVariation}
%%%%%%%%%%%%%%%%%%%%%%%%%%%%%%%
In this appendix we derive the first and second variations of the energy functional $\mathcal{E}$ defined in Eq.~(\ref{Eq.ConsEnergFunct}), discuss the relation of the second variation with the linearized system~(\ref{Eq:equationAB}), and show how this relation can be used to prove the absence of exponentially growing modes of ground state configurations having linear polarization. The methods discussed here build on previous arguments presented in Refs.~\cite{Nambo:2023yut, Nambo:2024gvs}.

Computing the first and second variations of $\mathcal{E}$ for arbitrary perturbations one obtains
\begin{subequations}
\label{Eq:Variation}
\begin{align}
\label{Eq:FirstVariation}
\delta\mathcal{E} &= 2\Re( \hat{\mathcal{H}}[\vec{\psi}]\vec{\psi},\delta\vec{\psi}), \\
\label{Eq:SecondVariation}
\delta^2\mathcal{E} &= 2\Re(\hat{\mathcal{H}}[\vec{\psi}]\vec{\psi},\delta^2 \vec{\psi})
 + 2(\hat{\mathcal{H}}[\vec{\psi}]\delta\vec{\psi},\delta\vec{\psi})\\ \nonumber
 &+ \frac{1}{2} (\delta n,\hat{K}[\delta n]) + \frac{\lambda_s}{2m_0^2}(\delta \vec{s},\delta \vec{s}),
\end{align}
\end{subequations}
where $\delta n := 2\Re(\vec{\psi}^*\cdot\delta \vec{\psi})$, $\delta \vec{s} := 2 \Re(- i \delta \vec{\psi}^* \times \vec{\psi})$ and we recall the definitions~(\ref{eq.def.H}) and~(\ref{Eq:Kdef}) of the formally self-adjoint operators $\hat{\mathcal{H}}[\vec{\psi}]$ and $\hat{K}$. Furthermore, $(\vec{\psi},\vec{\phi})=\int (\vec{\psi}^*\cdot\vec{\phi}) dV$ denotes the standard $L^2$-scalar product between $\vec{\psi}$ and $\vec{\phi}$. Together with Eq.~(\ref{integro-differential_equation}), the first variation implies that
\begin{equation}
\frac{d\mathcal{E}}{dt} 
 = 2\Re\left( \hat{\mathcal{H}}[\vec{\psi}]\vec{\psi}, \frac{1}{i}\hat{\mathcal{H}}[\vec{\psi}]\vec{\psi} \right) = 0,
\end{equation}
which shows that $\mathcal{E}$ is conserved, as expected.

Next, we evaluate $\delta^2\mathcal{E}$ at an equilibrium configuration [see Eq.~(\ref{eq.harmonic})], for which $\hat{\mathcal{H}}[\vec{\psi}] = \hat{E}\vec{\psi}$. For  stationary states $\hat{E}$ is proportional to the identity matrix, and we restrict the variation to the subset keeping the total particle number $N = (\vec{\psi},\vec{\psi})$ constant, such that $0 = \delta N = 2\Re(\vec{\psi},\delta\vec{\psi})$ and $0 = \delta^2 N = 2(\delta\vec{\psi},\delta\vec{\psi}) + 2\Re(\vec{\psi},\delta^2\vec{\psi})$. In contrast, for multi-frequency states, we need to keep fixed the quantity $\hat{Q}$ defined in Eq.~(\ref{eq.globalQ}), which yields
\begin{align}
0 &= \delta\hat{Q} = \int \left( \vec{\psi}^*\otimes \delta\vec{\psi} + \delta\vec{\psi}^*\otimes \vec{\psi} \right) dV,
\\
0 &= \delta^2\hat{Q} = 2\int \delta\vec{\psi}^*\otimes \delta\vec{\psi} dV
\nonumber\\
&+ \int \vec{\psi}^*\otimes \delta^2\vec{\psi} dV + \int \delta^2\vec{\psi}^*\otimes\vec{\psi} dV.
\end{align}
In either case one obtains the following expression for the second variation of the energy functional:
\begin{align}\label{Eq:SecondVariation}
\frac{1}{2}\delta^2\mathcal{E} &=
 (\vec{\sigma},(\hat{\mathcal{H}}^{(0)} - \hat{E})\vec{\sigma}) 
 + (\Re \vec{\sigma}^{(0)}\cdot\vec{\sigma}^*,\hat{K}[\Re \vec{\sigma}^{(0)}\cdot\vec{\sigma}^*]) 
\nonumber\\
 &+ \frac{\lambda_s}{m_0^2}(\Im \vec{\sigma}^{(0)}\times\vec{\sigma}^*,\Im\vec{\sigma}^{(0)}\times\vec{\sigma}^*),
\end{align}
where we have used the relation $\delta\vec{\psi} = e^{-i\hat{E} t}\vec{\sigma}$ which follows from Eq.~(\ref{eq:ansatzPert}).

In order to make further progress, we evaluate this formulae for the mode ansatz~(\ref{Eq:PertAnsatz}), which yields
\begin{equation}
\frac{1}{2}\delta^2\mathcal{E} = F(\vec{\mathcal{A}},\vec{\mathcal{B}}) e^{2\lambda t} + F(\vec{\mathcal{A}},\vec{\mathcal{B}})^* e^{2\lambda^* t} + K(\vec{\mathcal{A}},\vec{\mathcal{B}}) e^{2\lambda_R t},
\end{equation}
with
\begin{align}
F(\vec{\mathcal{A}},\vec{\mathcal{B}}) &:= \left( (\vec{\mathcal{A}} - \vec{\mathcal{B}})^*, (\hat{\mathcal{H}}^{(0)} - \hat{E})(\vec{\mathcal{A}} + \vec{\mathcal{B}})\right)
\nonumber\\
 &+ (\alpha^*,\hat{K}[\alpha]) - \frac{\lambda_s}{m_0^2}(\vec{\beta}^*,\vec{\beta} ),
\\
K(\vec{\mathcal{A}},\vec{\mathcal{B}}) &:= \left( (\vec{\mathcal{A}} + \vec{\mathcal{B}}), (\hat{\mathcal{H}}^{(0)} - \hat{E})(\vec{\mathcal{A}} + \vec{\mathcal{B}})\right)
\nonumber\\
 &+ \left( (\vec{\mathcal{A}} - \vec{\mathcal{B}}), (\hat{\mathcal{H}}^{(0)} - \hat{E} - i\frac{\lambda_s}{m_0^2}\vec{s}_0\times)(\vec{\mathcal{A}} - \vec{\mathcal{B}})\right)
\nonumber\\
 &+ 2(\alpha,\hat{K}[\alpha]) - \frac{2\lambda_s}{m_0^2}(\vec{\beta},\vec{\beta} ),
\end{align}
where we have abbreviated
\begin{align}
\alpha &:= \frac{1}{2}\left[ \vec{\sigma}^{(0)}\cdot(\vec{\mathcal{A}} - \vec{\mathcal{B}}) + {\vec{\sigma}^{(0)}}{}^*\cdot(\vec{\mathcal{A}} + \vec{\mathcal{B}}) \right],
\\
\vec{\beta} &:= \frac{1}{2}\left[ {\vec{\sigma}^{(0)}}{}^*\times(\vec{\mathcal{A}} + \vec{\mathcal{B}}) - \vec{\sigma}^{(0)}\times(\vec{\mathcal{A}} - \vec{\mathcal{B}}) \right].
\end{align}
After some calculations one can verify that the linearized equations~(\ref{Eq:equationAB}) imply that $F(\vec{\mathcal{A}},\vec{\mathcal{B}}) = 0$ and $K(\vec{\mathcal{A}},\vec{\mathcal{B}}) = 4i\lambda\Re(\vec{\mathcal{A}},\vec{\mathcal{B}})$, such that for solutions of the linearized equations,
\begin{equation}
\delta^2\mathcal{E} = 8i\lambda\Re(\vec{\mathcal{A}},\vec{\mathcal{B}}) e^{2\lambda_R t}.
\end{equation}
Since $\delta^2\mathcal{E}$ must be independent of $t$, it follows either that $\lambda_R = 0$ and $\delta^2\mathcal{E} = -8\lambda_I\Re(\vec{\mathcal{A}},\vec{\mathcal{B}})$ or that $\lambda_R \neq 0$ and $\delta^2\mathcal{E} = 0$. In particular, if $\delta^2\mathcal{E}$ is positive definite, only the first case can occur and there are no unstable modes.

In general, for a state which minimizes the energy, one can only guarantee that $\delta^2\mathcal{E}$ is positive {\it semi-definite}; thus the above arguments do not exclude the case of unstable modes with $\Re(\mathcal{A},\mathcal{B}) = 0$. However, we now prove that when $\lambda_0\geq 0$ and $\lambda_s\geq 0$ there are no such modes for  stationary states which minimize the energy functional $\mathcal{E}$ under the condition that $N$ is constant. As shown in Ref.~\cite{Nambo:2024hao}, these states have constant linear polarization (up to a symmetry transformation when $\lambda_s=0$), such that $\vec{\sigma}^{(0)}$ is real-valued and $\vec{s}_0 = 0$, which considerably simplifies the analysis. To prove the claim, we first notice that Eq.~(\ref{Eq:SecondVariation}) implies that
\begin{equation}
\frac{1}{2}\delta^2\mathcal{E} = (\vec{\sigma}_R,\hat{\mathcal{H}}_1\vec{\sigma}_R) +  (\vec{\sigma}_I,\hat{\mathcal{H}}_2\vec{\sigma}_I),
\end{equation}
with the formally self-adjoint operators $\hat{\mathcal{H}}_{1,2}$ defined by
\begin{align}
\hat{\mathcal{H}}_1\vec{\sigma} &:= (\hat{\mathcal{H}}^{(0)} - E)\vec{\sigma} + \hat{K}[\vec{\sigma}^{(0)}\cdot \vec{\sigma}]\vec{\sigma}^{(0)},
\\
\hat{\mathcal{H}}_2\vec{\sigma} &:= (\hat{\mathcal{H}}^{(0)} - E)\vec{\sigma}
 + \frac{\lambda_s}{m_0^2}[\vec{\sigma}^{(0)}\times \vec{\sigma}]\vec{\sigma}^{(0)}
\end{align}
when acting on a vector-valued function $\vec{\sigma}$.

For a ground state solution which minimizes the energy it follows that both $\hat{\mathcal{H}}_1$ and $\hat{\mathcal{H}}_2$ are positive semi-definite. On the other hand, Eqs.~(\ref{Eq:equationAB}) can be rewritten in the form $i\lambda\vec{\mathcal{A}} = \hat{\mathcal{H}}_2\vec{\mathcal{B}}$, $i\lambda\vec{\mathcal{B}} = \hat{\mathcal{H}}_1\vec{\mathcal{A}}$, which yields $i\lambda(\vec{\mathcal{B}},\vec{\mathcal{A}}) = (\vec{\mathcal{B}},\hat{\mathcal{H}}_2\vec{\mathcal{B}})\geq 0$ and $i\lambda(\vec{\mathcal{A}},\vec{\mathcal{B}}) = (\vec{\mathcal{A}},\hat{\mathcal{H}}_1\vec{\mathcal{A}})\geq 0$. This implies that $-\lambda^2|(\vec{\mathcal{A}},\vec{\mathcal{B}})|^2 \geq 0$. Hence, either $\lambda$ is purely imaginary or $\lambda_R\neq 0$ and $(\vec{\mathcal{A}},\vec{\mathcal{B}}) = 0$. In the second case it follows that $\vec{\mathcal{A}}$ must lie in the kernel of $\hat{\mathcal{H}}_1$ and $\vec{\mathcal{B}}$ in the kernel of $\hat{\mathcal{H}}_2$, which leads to  $\vec{\mathcal{A}} = \vec{\mathcal{B}} = 0$.

Unfortunately, we have not been able to extend this result to the case $\lambda_0\ge 0$ and $\lambda_s<0$, where the ground state is circularly polarized and the expressions become more involved.

%%%%%%%%%%%%%%%%%%%%%%%%%%%%%%%
\section{Equivalence  between the systems in Secs.~\ref{sec.linear.pol} and~\ref{sec.circular.pol} when $\lambda_s=0$} \label{Sec:equivalence}
%%%%%%%%%%%%%%%%%%%%%%%%%%%%%%%
In the absence of spin-spin self-interactions, all constant polarization states are related to each other by a global unitary transformation $\hat{U}$, which leaves the action~(\ref{eq.action.nonrel}) invariant when $\lambda_s=0$. In particular, linear and circular polarization vectors are related through $\hat{\epsilon}_z^{(+)}=\hat{U}\vec{e_x}$, with 
\begin{equation}
\hat{U} = \frac{1}{\sqrt{2}}
\left( \begin{array}{ccc}
1 & i & 0  \\
i & 1 & 0  \\ 
 0 & 0 & \sqrt{2}  \\
\end{array} \right).
\end{equation}
This unitary transformation maps the stationary state $\vec{\psi}(t,\vec{x})=e^{-iEt}\sigma^{(0)}(r)\vec{e}_x$ into $\vec{\psi}(t,\vec{x})=e^{-iEt}\sigma^{(0)}(r)\hat{\epsilon}_z^{(+)}$, and using Eqs.~(\ref{Eq:PertAnsatz}) and (\ref{Eq:CartesianPert}) it induces the following transformation in the perturbations,
\begin{subequations}\label{eqs.trans.circular.linear}
\begin{eqnarray}
A_{JM}^{x,c} &=& \frac{1}{\sqrt{2}}\left(A_{JM}^{x,l}+iB_{JM}^{y,l}\right),\\
B_{JM}^{x,c} &=& \frac{1}{\sqrt{2}}\left(B_{JM}^{x,l}+iA_{JM}^{y,l}\right),\\
A_{JM}^{y,c} &=& \frac{1}{\sqrt{2}}\left(A_{JM}^{y,l}+iB_{JM}^{x,l}\right),\\
B_{JM}^{y,c} &=& \frac{1}{\sqrt{2}}\left(B_{JM}^{y,l}+iA_{JM}^{x,l}\right),\\
A_{JM}^{z,c} &=& A_{JM}^{z,l},\\
B_{JM}^{z,c} &=& B_{JM}^{z,l},
\end{eqnarray}
\end{subequations}
where the superscripts $c$ and $l$ denote circular and linear polarization, respectively. If the transformation~(\ref{eqs.trans.circular.linear}) is applied to the equations  of Sec.~\ref{sec.circular.pol}, they reduce to those  of Sec~\ref{sec.linear.pol}.

%%%%%%%%%%%%%%%%%%%%%%%%%%%%%%%
\section{Numerical implementation of the linearized system}\label{Sec:numerical.study}
%%%%%%%%%%%%%%%%%%%%%%%%%%%%%%%
In this appendix, we present the details corresponding to the numerical implementation of the linear stability problem codified in our Eq.~(\ref{eq.general system}). Our methodology is based on previous works, see Refs.~\cite{Roque:2023sjl, Nambo:2023yut, Nambo:2024gvs}, thus here we only provide necessary details.

Linear mode-stability is determined by the properties of the eigenvalues $\lambda$ of the system~(\ref{eq.general system}), where the matrices $M_{JM}^{ij}$ are dependent on the
background solutions $(\vec\sigma^{(0)}, \vec u^{(0)})$ that are obtained as explained in Ref.~\cite{Nambo:2024hao}. To compute these eigenvalues, it is convenient to define the new rescaled functions $x_J(r):=r X_{JM}(r), y_J(r):=r Y_{JM}(r)$ and $z_J(r):=r Z_{JM}(r)$. Notice that, given that the equations are independent of the magnetic quantum number $M$, this label will be omitted in the following discussion. Introducing the rescaled functions in Eqs.~(\ref{eq.general system}) we obtain, after some algebra, a system with the general structure:
\begin{align}\label{eq.Lingeneral.system}
\left( \begin{array}{ccc}
 \mathcal{M}_{J}^{11} & \mathcal{M}_{J}^{12} & \mathcal{M}_{J}^{13}   \\[0.2cm]
 \mathcal{M}_{J}^{21} & \mathcal{M}_{J}^{22} & \mathcal{M}_{J}^{23}   \\[0.2cm] 
 \mathcal{M}_{J}^{31}  & \mathcal{M}_{J}^{32}  & \mathcal{M}_{J}^{33}  \\
\end{array} \right)\left( \begin{array}{c}
x_{J} \\[0.2cm]  y_{J} \\[0.2cm]  z_{J} \\ \end{array} \right)=-i\lambda \left( \begin{array}{c}
x_{J} \\[0.2cm]  y_{J} \\[0.2cm]  z_{J} \end{array} \right)\,,
\end{align}
where the matrices $\mathcal{M}_J^{ij}$ are defined for each polarization vector as follows:

\begin{widetext}
\begin{enumerate}
\item[\it{1.}]{\it Linear Polarization}
\begin{subequations}\label{eq.Num.matriz.linear}
\begin{align}
\mathcal{M}_{J}^{11} &= \left(\begin{array}{cc} 0 &\frac{d^2}{dr^2}+U^{\text{eff}}_J-\lambda_n\sigma^{(0)}{}^{2} \\%[0.2cm]
\frac{d^2}{dr^2}+U^{\text{eff}}_J-3\lambda_n \sigma^{(0)}{}^2-2\sigma^{(0)}\left(\frac{d^2}{dr^2}-\frac{J(J+1)}{r^2}\right)^{-1}[\sigma^{(0)}]& 0  
\end{array}\right),\\
\mathcal{M}_{J}^{12} &= 0, \\%[0.3cm]
\mathcal{M}_{J}^{21} &= 0,\\
\mathcal{M}_{J}^{22} &=
 \left(\begin{array}{cc}
 0 & \frac{d^2}{dr^2}+U^{\text{eff}}_J-(\lambda_n+2\lambda_s)\sigma^{(0)}{}^2 \\%[0.2cm]
 \frac{d^2}{dr^2}+U^{\text{eff}}_J-\lambda_n\sigma^{(0)}{}^2 & 0  
\end{array} \right),\\
\mathcal{M}_{J}^{33}&=\mathcal{M}_{J}^{22}.
\end{align}
\end{subequations}
%%%%

\item[\it{2.}]{\it Circular Polarization}
\begin{subequations}\label{eq.Num.matriz.circular}
\begin{align}
\mathcal{M}_{J}^{11} &= \left(\begin{array}{cc} 0 &\frac{d^2}{dr^2}+U^{\text{eff}}_J-\lambda_n\sigma^{(0)}{}^{2} \\
\frac{d^2}{dr^2}+U^{\text{eff}}_J-(2\lambda_n+\lambda_s)\sigma^{(0)}{}^2-\sigma^{(0)}\left(\frac{d^2}{dr^2}-\frac{J(J+1)}{r^2}\right)^{-1}[\sigma^{(0)}]& 0
\end{array}\right), \\
\mathcal{M}_{J}^{12} &= i\left(\begin{array}{cc} \lambda_s \sigma^{(0)}{}^{2} & 0 \\
0 & (\lambda_n+2\lambda_s)\sigma^{(0)}{}^2+\sigma^{(0)}\left(\frac{d^2}{dr^2}-\frac{J(J+1)}{r^2}\right)^{-1}[\sigma^{(0)}]
\end{array}\right), \\
\mathcal{M}_{J}^{22} &= \left(\mathcal{M}_{J}^{11}\right)^{T},\\
\mathcal{M}_{J}^{21} &= -i\left(\begin{array}{cc}  (\lambda_n+2\lambda_s)\sigma^{(0)}{}^2+\sigma^{(0)}\left(\frac{d^2}{dr^2}-\frac{J(J+1)}{r^2}\right)^{-1}[\sigma^{(0)}] & 0 \\
0 & \lambda_s \sigma^{(0)}{}^{2} \end{array}\right),\\
\mathcal{M}_{J}^{33}&= \left(\begin{array}{cc}
 0 & \frac{d^2}{dr^2}+U^{\text{eff}}_J-(\lambda_n+\lambda_s)\sigma^{(0)}{}^{2}\\
 \frac{d^2}{dr^2}+U^{\text{eff}}_J-(\lambda_n+\lambda_s)\sigma^{(0)}{}^{2} & 0  
\end{array} \right).
\end{align}
\end{subequations}

\item[\it{3.}]{\it Radial Polarization}
\begin{subequations}\label{eq.Num.matriz.circular}
\begin{align}
\mathcal{M}_{J}^{11} &= \left(\begin{array}{cc} 0 &\frac{d^2}{dr^2}+U^{\text{eff}}_J-\frac{2}{r^2}-\lambda_n\sigma^{(0)}{}^{2} \\
\frac{d^2}{dr^2}+U^{\text{eff}}_J-\frac{2}{r^2}-3\lambda_n\sigma^{(0)}{}^2-2\sigma^{(0)}\left(\frac{d^2}{dr^2}-\frac{J(J+1)}{r^2}\right)^{-1}[\sigma^{(0)}]& 0
\end{array}\right),\\
\mathcal{M}_{J}^{12} &= \displaystyle{\frac{2}{r^2}}\left(\begin{array}{c c c}
 0 & & \sqrt{J(J+1)}\\
 \sqrt{J(J+1)}& & 0
\end{array} \right), \\
\mathcal{M}_{J}^{21}&=\mathcal{M}_{J}^{12},\\
\mathcal{M}_{J}^{22}&=
 \left(\begin{array}{cc}
 0 & \frac{d^2}{dr^2}+U^{\text{eff}}_J-(\lambda_n+2\lambda_s)\sigma^{(0)}{}^{2}\\
 \frac{d^2}{dr^2}+U^{\text{eff}}_J-\lambda_n\sigma^{(0)}{}^{2} & 0  
\end{array} \right), \\ 
\mathcal{M}_{J}^{33}&=\mathcal{M}_{J}^{22}.
\end{align}
\end{subequations}

\item[\it{4.}]{\it Multi-frequency}
\begin{subequations}\label{eq.Num.matriz.multi_frequency}
\begin{align}
\mathcal{M}_{J}^{ii} &= \left( \begin{array}{cc}
 0 & \frac{d^2}{dr^2}+U^{\text{eff}}_{J,i}-\lambda_n\sum\limits_{j}\sigma_{j}^{(0)2} \\ \frac{d^2}{dr^2}+U^{\text{eff}}_{J,i}-\lambda_n\left(2\sigma_i^{(0)2}+
 \sum\limits_{j}\sigma_{j}^{(0)2}\right)
-2\sigma_i^{(0)}\left(\frac{d^2}{dr^2}-\frac{J(J+1)}{r^2}\right)^{-1}\left[\sigma_i^{(0)}\right] &  0 
\end{array} \right),\\
\mathcal{M}_{J}^{ij} &= -\left( \begin{array}{cc}
 0 & 0 \\2\lambda_n\sigma_{i}^{(0)}\sigma_{j}^{(0)}
+2\sigma_i^{(0)}\left(\frac{d^2}{dr^2}-\frac{J(J+1)}{r^2}\right)^{-1}\left[\sigma_j^{(0)}\right] &  0 
\end{array} \right),\quad i\neq j.
\end{align}
\end{subequations}
\end{enumerate}
\end{widetext}

Note that the subscript on the background quantities $(\vec\sigma^{(0)}, \vec u^{(0)})$ is omitted for linear, circular, and radial polarization to simplify the notation, as their respective components remain the same. We have introduced the effective potential $U^{\text{eff}}_{J,i}(r):= u_i^{(0)}(r) - J(J+1)/r^2$ and used the relations:
\begin{subequations}
\begin{align}
r\left(\hat{\mathcal{H}}^{(0)}-E\right)[r^{-1}] &= -\frac{d^2}{dr^2}-U^{\text{eff}}_J+\lambda_n\sigma^{(0)}{}^{2}\,,\\
rK_J[\sigma^{(0)}r^{-1}]&=\left(\frac{d^2}{dr^2}-\frac{J(J+1)}{r^2}\right)^{-1}[\sigma^{(0)}] \nonumber \\
&\quad+\lambda_n\sigma^{(0)}\,,
\end{align}
\end{subequations}
where the operator $\left(d^2/dr^2-J(J+1)/r^2\right)^{-1} = r\Delta_J^{-1} (r^{-1})$ denotes the inverse of $r \Delta_J(r^{-1})$ with homogeneous Dirichlet boundary conditions at $r=0$ and $r\to\infty$.

%%%%%%%%%%%%%%%%%%%%%%%%%%%%
\subsection{Boundary conditions:}
%%%%%%%%%%%%%%%%%%%%%%%%%%%%
To solve the system~(\ref{eq.Lingeneral.system}), we need two boundary conditions for each component of $x_J, y_J$, and $z_J$, which gives a total of $12$ boundary conditions. To determine them, one can study (heuristically) the dominant terms near the origin and infinity.

Using the fact that, near $r=0$, the background solution behaves as $\sigma^{(0)}\sim (1-\gamma)\sigma_{0}$ and $u^{(0)}\sim u_0$, one finds that for linear and circular polarization vectors the dominant terms at the center stem from the centrifugal term $J(J+1)/r^2$ in the effective potential $U^{\text{eff}}_J$, while for the radial polarization also the term $2/r^2$ contributes. Henceforth, near the origin, we can consider the system~(\ref{eq.Lingeneral.system}) as a set of differential equations of the form
\begin{align}
    Z'' -\frac{J(J+1)}{r^2}Z-\frac{2\gamma}{r^2}Z\approx0,
\end{align}
with $Z$ a column vector corresponding to the respective rescaled functions and $\gamma=0$ for linear and circular polarizations and $\gamma=1$ for radial polarization. Notice that the regular solution behaves as $Z(r)\sim r^{\frac{1}{2}+\sqrt{2\gamma+\left(\frac{2J+1}{2}\right)^2}}$ near $r=0$, which leads to the following boundary conditions at the origin:
\begin{subequations}\label{Eq.BounD.Orig}
\begin{align}
\lim\limits_{r \to 0} x^{(1)}(r)=0,\qquad \lim\limits_{r \to 0} x^{(2)}(r)=0, \\
\lim\limits_{r \to 0} y^{(1)}(r)=0,\qquad \lim\limits_{r \to 0} y^{(2)}(r)=0, \\
\lim\limits_{r \to 0} z^{(1)}(r)=0,\qquad \lim\limits_{r \to 0} z^{(2)}(r)=0.
\end{align}
\end{subequations}
The superscript $(1, 2)$ in these expressions indicates the first and second components of the corresponding eigenfields (e.g. for the radial polarization $x_J^{(1)}=r A^{r}_{JM}, x_J^{(2)}= r B^{r}_{JM}$).

In the asymptotic region, for every polarization, $\sigma^{(0)}$ decays exponentially to zero and $u^{(0)}$ approaches to the respective configuration energy eigenvalue $E$. This implies the reduction of the system~(\ref{eq.Lingeneral.system}) to differential equations of the form
\begin{align}
    Z''+E Z\approx 0 \,,
\end{align}
whose solution with finite total energy decays exponentially. Hence, the components of $(x_J, y_J, z_J)$ must vanish at infinity,
\begin{subequations} \label{Eq.BounD.Asin}
\begin{align}
\lim\limits_{r\to\infty} x^{(1)}(r)=0,\qquad \lim\limits_{r\to\infty} x^{(2)}(r)=0, \\
\lim\limits_{r\to\infty} y^{(1)}(r)=0,\qquad \lim\limits_{r\to\infty} y^{(2)}(r)=0, \\
\lim\limits_{r\to\infty} z^{(1)}(r)=0,\qquad \lim\limits_{r\to\infty} z^{(2)}(r)=0.
\end{align}
\end{subequations}

This implies that the system~(\ref{eq.Lingeneral.system}) can be solved by imposing the homogeneous Dirichlet boundary conditions given in Eqs.~(\ref{Eq.BounD.Orig}) and~(\ref{Eq.BounD.Asin}).

%%%%%%%%%%%%%%%%%%%%%%%%%%%%
\subsection{Discretization:}
%%%%%%%%%%%%%%%%%%%%%%%%%%%%
To determine the eigenvalues $\lambda$, we need to compute the spectrum of the matrix $(\mathcal{M}_J)^{ij}$. For this, we perform the discrete version of Eq.~(\ref{eq.Lingeneral.system}). Next, we provide the main details about the discretization procedure. We refer the reader to our previous works~\cite{Roque:2023sjl, Nambo:2023yut, Nambo:2024gvs} for a deeper exploration of this topic. 

The discretization of the system~(\ref{eq.Lingeneral.system}) is carried out in terms of Chebyshev polynomials using a standard spectral method. First, we map the radial variable, whose physical domain is $\mathsf{D}:=[0, r_\star]$ (truncated at a large radius $r_\star$), into the Chebyshev domain $\mathsf{D_{C}}:=[-1, 1]$ through the transformation $r=r_\star(x+1)/2$ with $x\in \mathsf{D_{C}}$ and corresponding to the Chebyshev points: $x_j=\cos{(j\pi/\mathsf{N})}$, with $j=0,1,\dots,\mathsf{N}$. Second, the derivative operator $d/dr$ is discretized using the $(\mathsf{N}+1)\times (\mathsf{N}+1)$ Chebyshev differentiation matrix as: $d/dr=(2/r_\star)\mathbb{D}_{\mathsf{N}}$ (in Chapter 6 of Ref.~\cite{trefethen2000spectral} the discrete operator $\mathbb{D}_{\mathsf{N}}$ is defined). The corresponding discretization of the second derivative operator $d^2/dr^2$ yields $(2/r_{\star})^2\mathbb{D}^2_{\mathsf{N}}$. Third, to impose the homogeneous Dirichlet boundary condition~(\ref{Eq.BounD.Orig},~\ref{Eq.BounD.Asin}) we use the procedure described in Chapter~7 of Ref.~\cite{trefethen2000spectral}, which involves striking the first and last rows and columns in the second derivative differentiation operator $\mathbb{D}_{\mathsf{N}}^{2}$, resulting in an $(\mathsf{N}-1)\times (\mathsf{N}-1)$ matrix $\mathbb{\tilde{D}}_{\mathsf{N}}$. Finally, the operator $\left(d^2/dr^2-J(J+1)/r^2\right)^{-1}$ is computed as the inverse of the matrix $\left[(2\mathbb{\tilde{D}}_{\mathsf{N}}/r_{\star})^{2}-\mathbb{J}\right]$ where the diagonal matrix $\mathbb{J}$ is defined below. 

Using everything previously mentioned, the system~(\ref{eq.Lingeneral.system}) is reduced to the finite-dimensional eigenvalue problem associated with a matrix $(\mathcal{M}_J)^{ij}$ of dimension $6(\mathsf{N}-1)\times 6(\mathsf{N}-1)$ and the vector
{\small
\begin{align}
&\begin{pmatrix} x_{J} \\ y_{J} \\ z_{J}\end{pmatrix}=\bigg
(x_J^{(1)}(r_1), \dots, x_J^{(1)}(r_{\mathsf{N}-1}), x_J^{(2)}(r_1),\dots, x_J^{(2)}(r_{\mathsf{N}-1}), \nonumber\\
& \quad\; y_J^{(1)}(r_1), \dots, y_J^{(1)}(r_{\mathsf{N}-1}), y_J^{(2)}(r_1), \dots, y_J^{(2)}(r_{\mathsf{N}-1}),\nonumber\\ 
& \quad\, z_J^{(1)}(r_1), \dots, z_J^{(1)}(r_{\mathsf{N}-1}), z_J^{(2)}(r_1), \dots, z_J^{(2)}(r_{\mathsf{N}-1})\bigg)^{T},
\end{align}
}%
corresponds to the discrete representation of the eigenfields $r(X_{JM}, Y_{JM}, Z_{JM})^T$. 

The matrix $(\mathcal{M}_J)^{ij}$ is a block matrix partitioned in blocks of smaller $2(\mathsf{N}-1)\times 2(\mathsf{N}-1)$ matrices. The blocks $\mathcal{M}_J$ involve the $(\mathsf{N}-1)\times (\mathsf{N}-1)$ matrices:
\begin{subequations}
\begin{align}
\Sigma_{0}&:=\textbf{diag}\left(\sigma^{(0)}(x_1), \sigma^{(0)}(x_2), \dots, \sigma^{(0)}(x_{\mathsf{N-1}})\right),\nonumber\\
U_{\text{eff}}&:=\textbf{diag}\left(u^{(0)}(x_1), u^{(0)}(x_2), \dots, u^{(0)}(x_{\mathsf{N-1}})\right)-\mathbb{J}, \nonumber\\
\mathbb{J} &:= \textbf{diag}\left(\frac{J(J+1)}{x_1^{2}}, \frac{J(J+1)}{x_2^{2}}, \dots, \frac{J(J+1)}{x_{\mathsf{N-1}}^{2}}\right), \nonumber
\end{align}
\end{subequations}
which are the respective discretizations of the background quantities $\sigma^{(0)}(r), U^{\text{eff}}_J(r)$, and the term $\frac{J(J+1)}{r^2}$. Note that in the multi-frequency case, $\Sigma_0$ and $U_{\text{eff}}$ have an additional index $i$ labeling the component of $\vec{\sigma}^{(0)}$ and $\vec{u}^{(0)}$.

Finally, once the discrete version of Eq.~(\ref{eq.Lingeneral.system}) is obtained, the finite-dimensional eigenvalue problem is solved using the SciPy library~\cite{2020SciPy-NMeth}. Our code is publicly available at~\cite{Roque_On_the_radial_2023}.

%%%%%%%%%%%%%%%%
\bibliography{ref.bib} 
%%%%%%%%%%%%%%%%

\end{document}